\documentclass[%
 aip,
 amsmath,amssymb,
 reprint,%
]{revtex4-1}

\usepackage{hyperref}
\hypersetup{%
	colorlinks=true, citecolor=blue, filecolor=blue, linkcolor=blue, urlcolor=blue}
\usepackage{braket}
\usepackage{natbib}

\usepackage{nomencl}

\usepackage{mathrsfs}
\usepackage{subfigure}

\usepackage{CJKutf8,CJKspace,CJKpunct}

\usepackage[dvipdfmx]{graphicx}
\usepackage{dcolumn}
\usepackage{bm}

\begin{document}

\title{Lattice Boltzmann modeling and simulation of forced-convection boiling \\ on a cylinder}

%
%
%
%
%
%

\author{Shimpei Saito}
\email{saito.shimpei@gmail.com}
\affiliation{Graduate School of Systems and Information Engineering, University of Tsukuba, Tsukuba 305-8573, Japan}

\author{Alessandro De Rosis}
\affiliation{Department of Mechanical, Aerospace and Civil Engineering, The University of Manchester, Manchester M13 9PL, UK}

\author{Linlin Fei}
\affiliation{Center for Combustion Energy; Key Laboratory for Thermal Science and Power Engineering of Ministry of Education, Department of Energy and Power Engineering, Tsinghua University, Beijing 100084, China}

\author{Kai Hong Luo}
\affiliation{Department of Mechanical Engineering, University College London, London WC1E 7JE, UK}

\author{Ken-ichi Ebihara}
\affiliation{Center for Computational Science \& e-Systems, Japan Atomic Energy Agency, Tokai 319-1195, Japan}

\author{Akiko Kaneko}
\author{Yutaka Abe}
\affiliation{Faculty of Engineering, Information and Systems, University of Tsukuba, Tsukuba 305-8573, Japan}


\begin{abstract}
When boiling occurs in a liquid flow field, the phenomenon is known as forced-convection boiling.
We numerically investigate such a boiling system on a cylinder in a flow at a saturated condition.
To deal with the complicated liquid-vapor phase-change phenomenon, we develop a numerical scheme based on the pseudopotential lattice Boltzmann method (LBM).
The collision stage is performed in the space of central moments (CMs) to enhance numerical stability for high Reynolds numbers.
The adopted forcing scheme, consistent with the CMs-based LBM, leads to a concise yet robust algorithm.
Furthermore, additional terms required to {ensure} thermodynamic consistency are derived in a CMs framework.
The effectiveness of the present scheme is successfully tested against a series of boiling processes, including nucleation, growth, and departure of a vapor bubble for Reynolds numbers varying between $30$ and $30000$.
Our CMs-based LBM can reproduce all the boiling regimes, {\it i.e.}, nucleate boiling, transition boiling, and film boiling, without any artificial input such as initial vapor phase.
We find that the typical boiling curve, also known as the Nukiyama curve, appears  even though the focused system is not the pool boiling but the forced-convection system.
Also, our simulations support experimental observations of intermittent direct solid-liquid contact even in the film-boiling regime.
Finally, we provide quantitative comparison with the semi-empirical correlations for the forced-convection film boiling on a cylinder on the $Nu$-$Ja$ diagram.
\end{abstract}

\maketitle

\section{Introduction}

\begin{figure}
	\centering
	\includegraphics[width=0.92\linewidth]{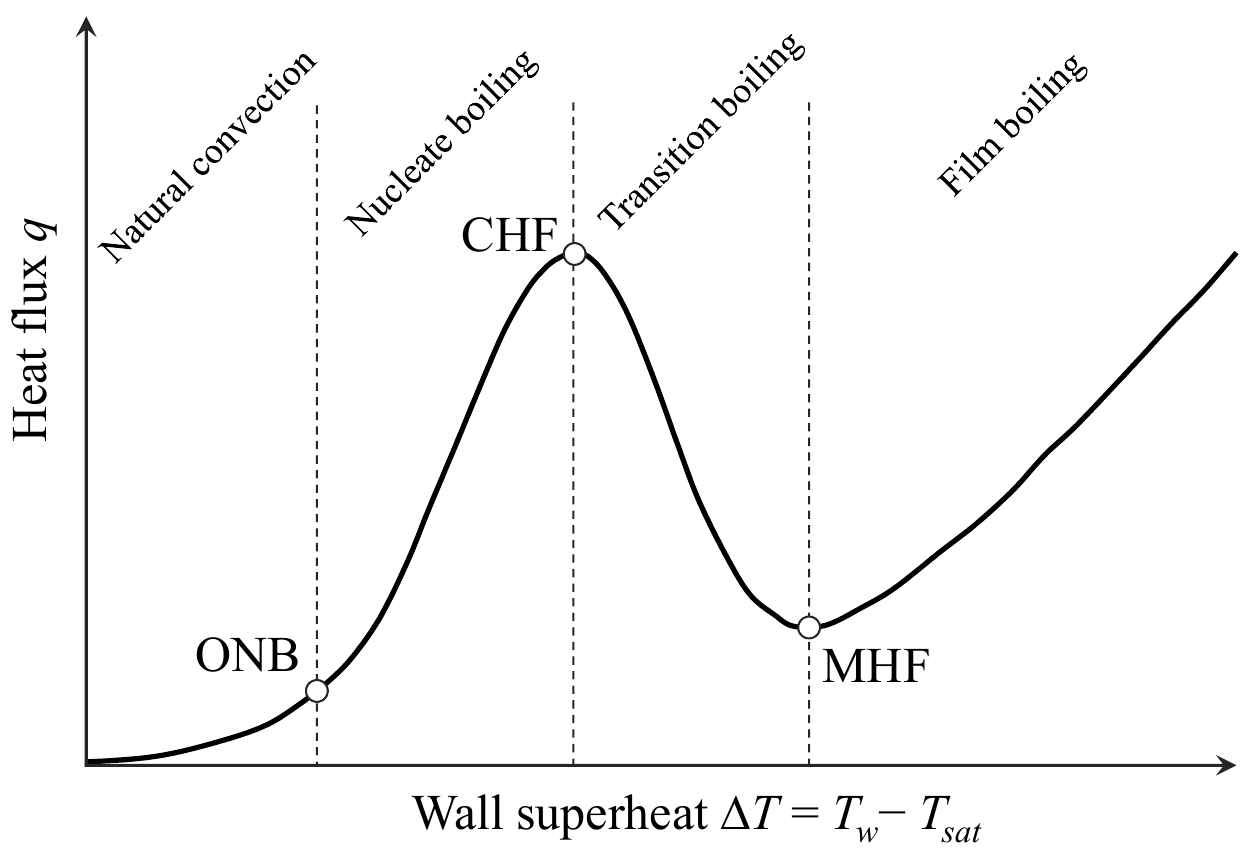}
	\caption{Typical boiling curve representing the relation between heat flux and wall superheat. ONB, CHF, and MHF denote onset of nucleate boiling, critical heat flux, and minimum film boiling, respectively.}
	\label{fig:typicalboilingcurve}
\end{figure}

Since the pioneering work of Nukiyama~\citep{Nukiyama1934}, boiling phenomena have long been investigated.
By referring to the so-called boiling curve (Fig.~\ref{fig:typicalboilingcurve}), three regimes can be identified: nucleate boiling, transition boiling, and film boiling with respect to surface superheat~\citep{Koizumi2017}. 
The passage from one regime to another is identified by salient characteristic points, {\it i.e.}, ONB (onset of nucleate boiling), CHF (critical heat flux), and MHF (minimum heat flux).
If the system has no condensation ability against the generated vapor, it is called saturated boiling. Otherwise, the system is regarded as subcooled boiling.
Pool boiling arises when vapor bubbles and heated liquid move only due to gravity, whereas the presence of other forces promotes the so-called forced-convection boiling.

\begin{figure}
	\centering
	\includegraphics[width=0.90\linewidth]{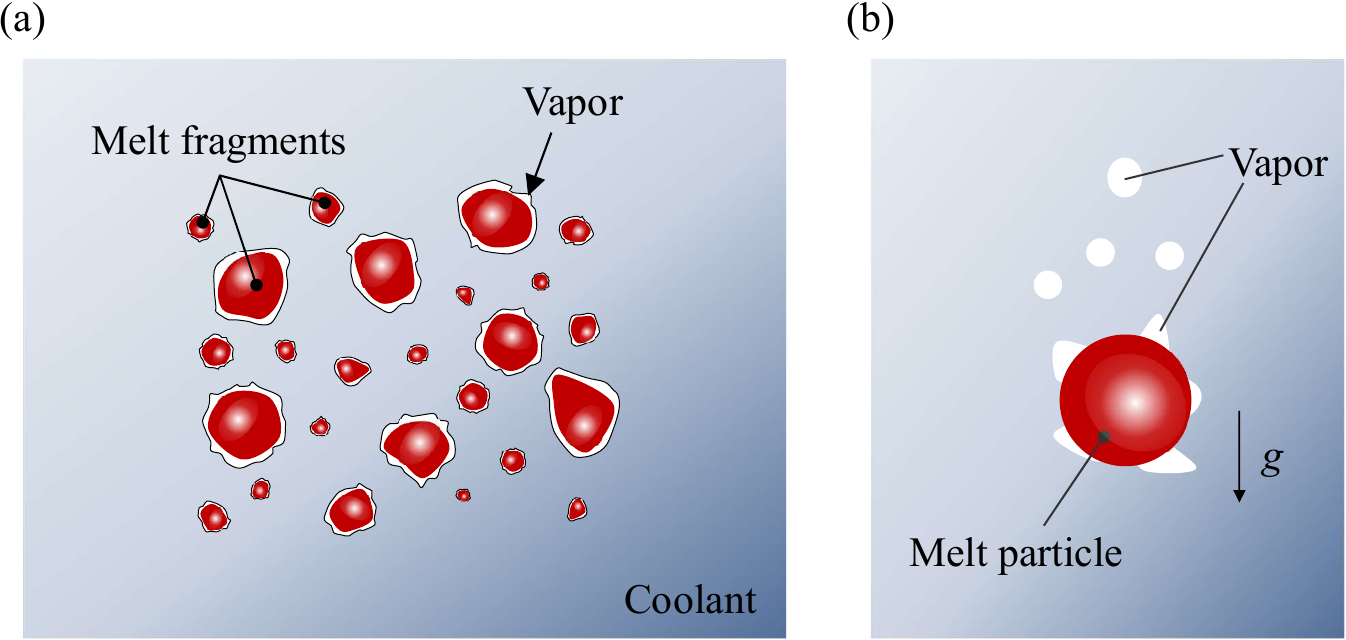}
	\caption{Schematic diagram of the possible FCI scenario: 
		(a) Melt fragments distributed in coolant, (b) A simplified situation with a single melt particle in a coolant.}
	\label{fig:fci}
\end{figure}

Investigating heat transfer coefficient (HTC) of forced-convection boiling on a sphere has been a critical topic, especially in the nuclear safety field.
That is because the HTC between a high-temperature melt particle and liquid plays a key role in modeling fuel-coolant interactions (FCIs) during any accidental event of nuclear power plants~\citep{Kobayasi1965,Dhir1978,Epstein1980,Liu1994}.
A schematic diagram of the possible FCI scenario is shown in Fig.~\ref{fig:fci}(a), where melt fragments are generated from the ejected melt jets penetrating the coolant.
Since the interaction between a cluster of fragments and coolant would be quite difficult to be modeled, the one between a single melt particle and liquid, as in Fig.~\ref{fig:fci}(b), is often considered in melt-coolability evaluation.
To this end, the HTC correlation of the forced-convection boiling, which is generally modeled within the film-boiling regime, is required.
One can find that the forced-convection boiling on a sphere or cylinder simulates the situation in Fig.~\ref{fig:fci}(b). 
The choice of such HTC correlations would lead to large uncertainties of melt-particle coolability. However, there has been little discussion about selecting an appropriate HTC correlation.

Starting from Bromley's study in chemical engineering~\citep{Bromley1950}, film-boiling heat transfer has attracted much attention.
Bromley {\it et al.}~\citep{Bromley1953} experimentally and analytically studied forced-convection film boiling on a cylinder at saturation temperature.
They also derived an appropriate theoretical analysis under steady-state assumptions.
Similar analyses for the cases of flow past a sphere were reported later by Kobayasi~\citep{Kobayasi1965} and Witte~\citep{Witte1968}.
Ito {\it et al.}~\cite{Ito1981} analyzed the phenomena with various liquid media as water, ethanol, and hexane, based on the two-phase boundary-layer theory.
Dhir and Purohit~\citep{Dhir1978} studied the pool and forced-convection film boiling on metal spheres under subcooled conditions from 0 to 50 K. 
Epstein and Hauser~\citep{Epstein1980} derived a semi-empirical correlation by solving conservation equations for vapor and liquid boundary layers at the stagnation point.

Liu and Theofanous~\citep{Liu1994} reviewed the existing studies of the forced-convection film boiling on a sphere or cylinder, in which they classified the film-boiling HTC correlations into two modes as Mode 1 and Mode 2.
In the dimensionless space, the film-boiling HTC is expressed as the Nusselt number:
\begin{equation}
{Nu} = \frac{h D}{\lambda_v},
\end{equation}
and is described in terms of the liquid-vapor density ratio $\gamma$, viscosity ratio $\eta$, liquid Reynolds number $Re$, Prandtl number $Pr$, and Jacob number ${Ja}$ as follows:
\begin{align}
\gamma &= \frac{\rho_l}{\rho_v}\\
\eta &= \frac{\nu_l}{\nu_v}\\
Re & = \frac{u_l D}{\nu_l}\\
Pr & = \frac{\nu_v}{\alpha_v}\\
{Ja} &= \frac{c_{p,v} \Delta T}{h_{fg}},
\end{align}
where $h$ is the heat transfer coefficient, $D$ is the characteristic length (typically, diameter of sphere or cylinder), $\lambda$ is the thermal conductivity,  $\rho$ is the density, $\nu$ is the kinematic viscosity, $u$ is the flow velocity, $\alpha$ is the thermal diffusivity, $c_{p}$ is the specific heat at constant pressure, $\Delta T$ is the superheat degree, and $h_{fg}$ is the latent heat of vaporization. 
The subscript $l$ and $v$ denote the liquid and the vapor phases, respectively.
Now let $C$ denote a given constant.
With dimensionless form, Mode 1 correlation~\citep{Bromley1953,Witte1968} can be written as
\begin{equation} \label{eq:mode1}
{Nu} = C
\, \eta^{1/2} 
\, Re^{1/2} 
\, Pr^{1/2} 
\, {Ja}^{-1/2}.
\end{equation}	
Mode 2 correlation~\citep{Epstein1980,Ito1981,Liu1994}, on the other hand, can be written as
\begin{equation} \label{eq:Mode2}
{Nu} = C
\, \gamma^{1/4} 
\, \eta^{1/2}
\, Re^{1/2} 
\, Pr^{1/4}
\, {Ja}^{-1/4}.
\end{equation}
Both types of correlations state that ${Nu} \propto \eta^{1/2} Re^{1/2} $, but have different dependencies on $\gamma$, $\eta$, $Pr$, and ${Ja}$.
Such correlations are usually validated with experimental data, but the experimental setup and conditions generally include unexpected uncertainties. Thus, numerical simulations would be an effective approach to further understanding the heat transfer characteristics of the phenomena.

Due to its important role in elucidating the mechanism and heat transfer characteristics, numerical simulations of boiling phenomena have been carried out since late 1990s~\citep{Son1997,Juric1998,Welch2000,Yazdani2016,Sato2017}.
The first attempt was made by \citet{Son1997}, who studied the evolution of the liquid-vapor interface during saturated film boiling with a level-set method.
\citet{Juric1998} extended a front-tracking method to simulate horizontal film boiling by adding a source term to the continuity equation.
\citet{Welch2000} proposed a volume-of-fluid based method to simulate horizontal film boiling.
Since then, a lot of numerical studies have been conducted to investigate boiling phenomena (see {\it e.g.}, Refs.~\citep{Kunugi2012, Dhir2013, Cheng2014, Kharangate2017} for further details).
{Recently, \citet{Sato2017} proposed a numerical simulation method
	to model nucleate pool boiling involving multiple nucleation sites based on a sharp-interface, phase-change solver.}
However, most of these methods assume an initial vapor phase as an {\it artificial input}. Therefore, they cannot simulate nucleation in boiling phenomena.

In recent years, as a different approach from the aforementioned Navier--Stokes-based method, several kinds of lattice Boltzmann methods (LBMs) for multiphase flows have been applied to simulating liquid-vapor phase-change phenomena.
Multiphase LBMs can be divided into four categories, namely color-gradient~\citep{Gunstensen1991,Grunau1993}, pseudopotential~\citep{Shan1993,Shan1994}, free-energy~\citep{Swift1995,Swift1996}, and mean-field~\citep{He1999} models.
This is not an exhaustive classification; for instance, the latter two model types are sometimes called phase-field models since the Cahn--Hiliard (or similar) interface tracking equations can be derived from them.
The multiphase LBMs have been successfully applied to a variety of complex multiphase-flow problems, such as liquid drop~\citep{Kang2005} and bubble~\citep{Frank2005} dynamics, flow through porous media~\citep{Parmigiani2011}, wetting-drying process~\citep{Liu2014}, liquid-jet breakup~\citep{Saito2017}, etc.
Among the multiphase LBMs, the pseudopotential and the phase-field models are generally used in phase-change simulations~\citep{Li2016}.
In most of the phase-field models, an interface-tracking equation is solved to capture the liquid-vapor interface and a source term is incorporated into the continuity or the Cahn--Hiliard equations to define the phase-change rate.
This implies that the rate of the liquid-vapor phase change is an {\it artificial input}.

In contrast, the pseudopotential model has no such a limitation.
The most distinct feature of this model is that the phase separation is achieved via an interparticle potential.
In the case considering temperature change, the liquid-vapor phase change is driven by the equation of state (EOS). 
Hence no artificial phase-change terms need to be added to the temperature equation~\citep{Li2016}.
After the first attempt by Zhang and Chen~\citep{Zhang2003}, many researchers proposed the LBM models to deal with the phase-change phenomena~\citep{albernaz2017, qin2019}, and succeeded in simulating boiling process without artificial terms~\citep{Hazi2009,Markus2012a,Markus2012b,Biferale2012,Gong2013,Mu2017,Li2020,Fei2020}.
{Focusing on the mesoscopic features of the LBM, \citet{Mu2017} investigated nucleate boiling performance in cavities with varied geometries.}
{\citet{Yu2020} investigated the boundary schemes in LBM simulations of boiling heat transfer involving curved surfaces.}
{Li {\it et al.}~\citep{Li2020} showed theoretically and numerically how boiling occurs in pseudopotential LBM simulations.}

Most of the aforementioned models have the common feature that a thermal lattice Boltzmann equation (LBE) with a temperature distribution function is used to mimic the macroscopic energy equation.
However, the forcing term included in the thermal LBEs introduces a spurious term into the macroscopic temperature equation~\citep{Li2015}, and such a forcing term leads to significant errors in the simulation of thermal flows~\citep{Li2014}.
Furthermore, in temperature-based thermal LBEs, another error term proportional to $\nabla \cdot (T\nabla \rho / \rho)$ appears in the macroscopic temperature equation, that can be seen in Refs.~\citep{Li2008,Chai2013}.   
This term yields considerable errors in multiphase flows with density varying.
According to Li {\it et al.}~\citep{Li2015}, solving the temperature equation using the classical finite-difference method enables us to be free from the problems.
{It should be mentioned that \citet{Zhang2017} have derived the source term so that the thermal LBE (or dual distribution function) approach also recovers the correct temperature equation.
Based on the theoretical analysis, Li {\it et al.}~\citep{Li2017} have eliminated the error terms and the discrete effect of the source term in the thermal LBE framework.}

Besides, the original pseudopotential model with the Bhatnagar--Gross--Krook (BGK)~\citep{Bhatnagar1954} approximation generally suffers from numerical instability under high-Re (low-viscosity) conditions.
One way to overcome this issue is to modify the collision operator~\citep{Luo2011,Coreixas2019}.
To cope with this problem, Lycett-Brown and Luo~\citep{Lycett-Brown2014a} introduced the cascaded scheme~\citep{Geier2006} into a pseudopotential model to enhance numerical stability.
Later, Lycett-Brown and Luo~\citep{Lycett-Brown2016} updated their multiphase LBM model with their original forcing scheme~\citep{Lycett-Brown2015}.
The cascaded collision operator relaxes the so-called central moments (CMs)~\citep{Geier2006,Geier2017,DeRosis2016,DeRosis2017, DeRosis2019}, instead of raw moments used in the multiple-relaxation-time (MRT) schemes~\citep{dHumieres1992,Lallemand2000}.
Undoubtedly, the CMs-based LBM drastically outperforms both BGK and MRT in terms of stability for high-Re multiphase flow~\citep{Lycett-Brown2014a,Lycett-Brown2016,Lycett-Brown2014b,Saito2018,DeRosis2019}.
{
	On the CMs-based thermal LB model, the recent works of Li {\it et al.}~\citep{Li2019b} and Shan~\citep{Shan2019} gave a concise and comprehensive theoretical framework.
}

In this paper, aiming at numerically investigating the forced-convection boiling on a cylinder up to high Reynolds numbers, we develop a numerical scheme based on the LBM.
The pseudopotential model and its forcing are formulated in the framework of nonorthogonal CMs.
The flow and energy equations are coupled through the EOS to represent the liquid-vapor phase change.
The developed numerical scheme is applied to the forced-convection boiling on a cylinder.

The rest of the paper is organized as follows.
Sec.~\ref{sec:formulation} describes formulation of the CMs-based pseudopotential LBM model with the consistent forcing scheme, and derivation of the macroscopic energy equation to be solved with finite-difference method.
Sec.~\ref{sec:tests} verifies the present scheme through {several} numerical tests, {including steady and unsteady problems}.
Sec.~\ref{sec:simulation} presents simulations of forced-convection boiling on a cylindrical body at $Re = 30$ and $30000$.
Obtained HTCs are compared with several semi-empirical correlations.
Finally, Sec.~\ref{sec:conclusions} concludes this paper.

\section{Methodology \label{sec:formulation}}


\subsection{Central-moments-based lattice Boltzmann equation}
Let us consider an Eulerian basis $\boldsymbol{x}=[x,y]$ and the D2Q9 velocity space~\citep{SucciBook}. 
The LBE predicts the space and time evolution of the particle distribution functions $\ket{f_{i}} = [ f_0,\, f_1,\, \dots,\, f_8 ]^\top$ colliding and streaming on a fixed square grid along the generic link $i=0 \ldots 8$ with lattice velocity $\boldsymbol{c}_i = [\ket{c_{ix}},\,\ket{c_{iy}}]$, where
\begin{equation}
\begin{split}
\ket{c_{ix}} = &[0,\,   1,\, -1,\,  0,\,  0,\,    1,\,  -1,\,   1,\,  -1]^\top, \\
\ket{c_{iy}} = &[0,\,   0,\,  0,\,  1,\, -1,\,    1,\,  -1,\,  -1,\,   1]^\top. 
\label{eq:velVec}
\end{split}
\end{equation}
Let us employ the symbols $\ket{\cdot}$ and $\top$ to denote a column vector and the transpose operator, respectively. 
The LBE with the forcing term can be generally expressed as~\citep{McCracken2005,Fei2017}
\begin{equation}
	\begin{split} \label{lbe}
	\ket{f_i(\boldsymbol{x}+\delta_t \boldsymbol{c}_i, t+\delta_t)} 
		=& \ket{f_i(\boldsymbol{x}, t)} 
		+ {\bm \Lambda} [ \ket{f_i^{\mathrm{eq}}(\boldsymbol{x}, t)} 
		-\ket{f_i(\boldsymbol{x}, t)} ] \\
		&+ \delta_t (\bm{I} - {\bm \Lambda}/2 ) \ket{F_i(\boldsymbol{x}, t)},
	\end{split}
\end{equation}
where  $ f_i^{\mathrm{eq}}$ is the discrete local equilibrium, and the time step is set to $\delta_t=1$.
When the collision matrix ${\bm \Lambda}$ has only one relaxation frequency as $\omega \bm{I}$, (\ref{lbe}) describes the so-called BGK LBE.
As usual, the LBE can be divided into two steps, \textit{i.e.}, collision
\begin{equation}
\begin{split}\label{collision}
\ket{f_i^{\star}(\boldsymbol{x}, t)} 
=& \ket{f_i(\boldsymbol{x}, t)} 
+ {\bm \Lambda} [ \ket{f_i^{\mathrm{eq}}(\boldsymbol{x}, t)} -\ket{f_i(\boldsymbol{x}, t)} ] \\
&+ \delta_t (\bm{I}- {\bm \Lambda}/2 ) \ket{F_i(\boldsymbol{x}, t)},
\end{split}
\end{equation}
and streaming
\begin{equation}\label{streaming}
f_i(\boldsymbol{x}+\delta_t \boldsymbol{c}_i, t+\delta_t) = f_i^{\star}(\boldsymbol{x}, t),
\end{equation}
where the superscript $\star$ denotes post-collision quantities here and henceforth. 
The dependence on the space and the time will be implicitly assumed in the rest of this section. 
The term $F_i$ accounts for external body forces $\boldsymbol{F} = [F_x, F_y]$ and its role will be elucidated later. The fluid density $\rho$ and velocity $\boldsymbol{u} = [u_x, u_y]$ are computed as 
\begin{equation}
\rho = \sum_i{f_i},~ \rho \boldsymbol{u} = \sum_i{f_i \boldsymbol{c}_i} + \frac{\boldsymbol{F}}{2}\delta_t,
\end{equation}	
respectively. 
Following the works by Malaspinas~\citep{malaspinas2015increasing} and Coreixas~\citep{coreixas2017recursive,COREIXAS_PhD_2018}, the equilibrium distribution can be expanded into a basis of Hermite polynomials $\bm{\mathcal{H}}^{(n)}$ as
\begin{equation} 
\begin{split} \label{eq:equi}
f_i^{\mathrm{eq}} = w_i \rho &\left[ 1+ \frac{{\bm c}_i \cdot \boldsymbol{u}}{c_s^2} + \frac{1}{2 c_s^4} \bm{\mathcal{H}}_i^{(2)} : \boldsymbol{u}\boldsymbol{u} \right. \\
& \left. +\frac{1}{2 c_s^6} \left( \bm{\mathcal{H}}_{ixxy}^{(3)} u_x^2 u_y + \bm{\mathcal{H}}_{ixyy}^{(3)} u_x u_y^2 \right) \right. \\
& \left.  +\frac{1}{4 c_s^8} \bm{\mathcal{H}}_{ixxyy}^{(4)}u_x^2 u_y^2 \right],
\end{split}
\end{equation}
with $w_0=4/9$, $w_{1 \ldots 4} = 1/9$, $w_{5 \ldots 8}=1/36$ and $c_s=1/\sqrt{3}$ is the lattice sound speed. Notice that the maximum order of the expansion is equal to four in the D2Q9 space. Moreover, the model recovers the classical second-order truncated equilibrium when $ \bm{\mathcal{H}}^{(3)}$ and $ \bm{\mathcal{H}}^{(4)}$ are neglected.

The pivotal idea to design any CMs-based collision operator is to shift the lattice directions by the local fluid velocity~\citep{Geier2006}. 
Therefore, it is possible to define $\displaystyle \bar{\bm{c}}_i=[\ket{\bar{c}_{ix}} ,\, \ket{\bar{c}_{iy}}]$, where 
\begin{equation}
\begin{split}
\ket{\bar{c}_{ix}} = & \ket{c_{ix}- u_x}, \\
\ket{\bar{c}_{iy}} = & \ket{c_{iy}- u_y}.
\end{split}
\end{equation}
Then, one must choose a suitable basis of moments. 
Let us adopt the non-orthogonal basis~\citep{DeRosis2016} as the following matrix form:
\begin{equation} \label{eq:matrix}
{\mathrm T} = 
\left [
\begin{array}{c}
\bra{|\boldsymbol{c}_i|^0} \\
\bra{\bar{c}_{ix}} \\
\bra{\bar{c}_{iy}} \\
\bra{\bar{c}_{ix}^2+ \bar{c}_{iy}^2} \\
\bra{\bar{c}_{ix}^2- \bar{c}_{iy}^2} \\
\bra{\bar{c}_{ix} \bar{c}_{iy}}  \\
\bra{\bar{c}_{ix}^2 \bar{c}_{iy}} \\
\bra{\bar{c}_{ix} \bar{c}_{iy}^2}  \\
\bra{\bar{c}_{ix}^2 \bar{c}_{iy}^2} 
\end{array}
\right] ,	
\end{equation}	
where $\bra{\cdot}$ denotes the raw vector.
Using equation (\ref{eq:matrix}), the collision matrix can be written as ${\bm \Lambda} = \bm{T}^{-1} \bm{K} \bm{T}$, where $\bm{K}$ is the relaxation matrix specified later.
Let us collect pre-collision, equilibrium and post-collision CMs as
\begin{equation}
\begin{split}
\ket{k_i} =& \left[ k_0,\, \ldots, \, k_i,\, \ldots, \, k_{8}   \right]^\top, \\
\ket{k_i^{\mathrm{eq}}} =& \left[ k_0^{\mathrm{eq}},\, \ldots, \, k_i^{\mathrm{eq}},\, \ldots, \, k_{8}^{\mathrm{eq}}   \right]^\top, \\
\ket{k_i^{\star}} =& \left[ k_0^{\star},\, \ldots, \, k_i^{\star},\, \ldots, \, k_{8}^{\star}   \right]^\top,
\end{split}
\end{equation}
respectively. 
The first two quantities are evaluated by applying the matrix ${\bm{T}}$ to the corresponding distribution, that is
\begin{equation}
\ket{k_i} 					  =  {\bm{T}} \ket{f_i}, ~	\ket{k_i^{\mathrm{eq}}}  =  {\bm{T}} \ket{f_i^{\mathrm{eq}}} ,
\end{equation}
where $\ket{f_i^{\mathrm{eq}}} = [f_0^{\mathrm{eq}} ,\, \ldots f_i^{\mathrm{eq}}, \, \ldots f_{8}^{\mathrm{eq}} ]^\top$. 
By adopting $n=4$ in the Hermite polynomials, equilibrium CMs can be computed as
\begin{equation}
k_0^{\mathrm{eq}} = \rho,~ k_3^{\mathrm{eq}} = \frac{2}{3}\rho,~ k_8^{\mathrm{eq}} = \frac{1}{9}\rho,
\end{equation}
with $k_{1,2}^{\mathrm{eq}} = k_{4\dots7}^{\mathrm{eq}} = 0$.
Notably, only three equilibrium CMs assume values different from zero. 
It is of interest to notice that the discrete equilibrium CMs have the same form of the continuous counterparts when the full set of Hermite polynomials is considered. 
The post-collision CMs can be written as
\begin{equation}
\begin{split}\label{eq:posColl0}
\ket{k_i^{\star}} 
=& \left( \bm{I}- \bm{K} \right) \bm{T} \ket{f_i} 
+ \bm{K}  \bm{T} \ket{f_i^{\mathrm{eq}}} + \left( \bm{I}
- \frac{\bm{K}}{2} \right) \bm{T} \ket{F_i},  \\
=& \left( \bm{I}- \bm{K} \right)  \ket{k_i} 
+ \bm{K}   \ket{k_i^{\mathrm{eq}}} + \left( \bm{I}
- \frac{\bm{K}}{2} \right)  \ket{R_i},
\end{split}
\end{equation}
where 
\begin{equation}
\bm{K} = \mathrm{diag} [1,\,1,\,1,\,1, \, \omega,\, \omega,\,1,\,1, \, 1  ],
\end{equation}
is a $9 \times 9$ relaxation matrix with $\displaystyle \omega = \left( \frac{\nu}{c_s^2 \delta_t}+\frac{1}{2}   \right)^{-1}$, $\nu$ being the fluid kinematic viscosity.
The CMs of the discrete force term $\ket{R_i} = \bm{T}\ket{F_i}$ are computed as follows.

Now, let us define the discrete forcing term $F_i$. 
Specifically, we employ the expression adopted by Huang {\it et al.}~\citep{Huang2018}:
\begin{equation}
\begin{split}
F_i (\boldsymbol{u}) = w_i & \left( \frac{\boldsymbol{F}}{c_s} \cdot  \bm{\mathcal{H}}^{(1)} 
+ \frac{[\boldsymbol{F} \boldsymbol{u}]}{2 c_s^2}  \cdot \bm{\mathcal{H}}^{(2)} \right. \\
& \left. + \frac{[\boldsymbol{F} \boldsymbol{u} \boldsymbol{u}]}{6 c_s^3}  \cdot \bm{\mathcal{H}}_{[xyy],[xxy]}^{(3)} 
+ \frac{[\boldsymbol{F} \boldsymbol{u} \boldsymbol{u} \boldsymbol{u}]}{24 c_s^4}  \cdot  \bm{\mathcal{H}}_{[xxyy]}^{(4)} \right),
\label{eq:forcingTerm}
\end{split}
\end{equation}
where the square bracket in Hermite coefficient denotes permutations ({\it e.g.}, $[\boldsymbol{F} \boldsymbol{u}\boldsymbol{u}]= \boldsymbol{F} \boldsymbol{u}\boldsymbol{u}+\boldsymbol{u}\boldsymbol{F} \boldsymbol{u}+\boldsymbol{u}\boldsymbol{u}\boldsymbol{F}$). 
Notice that the popular formula proposed by Guo {\it et al.}~\citep{Guo2002} is recovered if $\bm{\mathcal{H}}^{(3)}$ and $\bm{\mathcal{H}}^{(4)}$ are neglected. 
The CMs of the present discrete force term can be computed as 
\begin{equation}
\ket{R_i} = \bm{T} \ket{F_i },
\end{equation}
and read as follows:
\begin{equation}
R_1 = F_x,~ R_2 = F_y,~ R_6 = \frac{1}{3}F_y,~ R_7 = \frac{1}{3}F_x,
\label{eq:force1}
\end{equation}	
with $R_0 = R_{3\dots5} = R_8 = 0$.
{
	Note that both Eqs.~(\ref{eq:equi}) and (\ref{eq:forcingTerm}) are subsets of the more general expressions given in Shan {\it et al.}~\citep{Shan2006b}.
}

Some considerations should be drawn regarding these results. 
The same expressions of $\ket{R_i}$ can be achieved when the continuous Maxwell--Boltzmann distribution are considered~\citep{Premnath2009, Fei2017}. 
This is consistent with a recent work of De Rosis and Luo~\citep{DeRosisHermite}, where it is argued that the CMs of the discrete distribution collapse into the continuous counterpart when the full set of Hermite polynomials is considered. 
Moreover, $\ket{R_i}$ can be achieved by disregarding the velocity-dependent terms in Eq.~(15) in Ref.~\citep{DeRosis2017c}. 
Again, this is consistent with De Rosis and Luo~\citep{DeRosisHermite}, where the velocity terms vanish when the maximum $\bm{\mathcal{H}}$ is adopted to construct CMs. 
Finally, the present findings are different from and simpler than those in Huang {\it et al.}~\citep{Huang2018} [see Eq.~(9)] due to the adoption of a different basis. 

In the following, we report the post-collision CMs:
\begin{equation}
\begin{split}
k_0^{\star} = \rho,~ k_1^{\star} = \frac{1}{2}F_x,~ 	k_2^{\star} = \frac{1}{2}F_y,~ k_3^{\star} = \frac{2}{3}\rho, \\
k_4^{\star} = (1-\omega) k_4,~ 	k_5^{\star} = (1-\omega) k_5, \\
k_6^{\star} = \frac{1}{6}F_y,~ k_7^{\star} = \frac{1}{6}F_x,~ 	k_8^{\star} = \frac{1}{9}\rho.
\label{eq:postColl}
\end{split}
\end{equation}
One can immediately appreciate that the present scheme is highly intelligible and the resultant algorithm is very concise. 
Then, the post-collision populations are reconstructed as
\begin{equation}\label{system}
\ket{f_i^{\star}} 
= \bm{T}^{-1} \ket{k_i^{\star} },
\end{equation}
that are eventually streamed. 
For practical implementation, it is easier to 
replace the above ``one-step" reconstruction by the ``two-step" reconstruction. 
For more details, the interested readers are kindly directed to Fei and Luo~\citep{Fei2017}. 

\subsection{Pseudopotential and thermodynamic consistency}
In the pseudopotential model, the interaction force to mimic the molecular interactions plays as important role in phase separation, {which is given by~\citep{Shan1993}}
\begin{equation}
\boldsymbol{F}_m(\boldsymbol{x}) = -\mathscr{G} c_s^2 \, \psi(\boldsymbol{x}) \sum_{i=0}^N{w(|\boldsymbol{c}_i|^2) \, \psi(\boldsymbol{x}+ \boldsymbol{c}_i) \, \boldsymbol{c}_i},
\label{eq:interForce}
\end{equation}	
where $\psi$ is the interaction potential, $\mathscr{G}$ is the interaction strength, and $w(|\boldsymbol{c}_i|^2)$ are the weights.
The number of discrete velocities $N$ used in the force calculation need not to be equal to the number of lattice velocities. 
We consider the case of $N=24$ (pseudo D2Q25 lattice) for calculation of the interaction force. 
{
	The weights can be defined as $w(0) = 247/140$, $w(1) = 4/21$, $w(2) = 4/45$, $w(4) = 1/60$, $w(5) = 2/315$, and $w(8) = 1/5040$~\citep{Shan2006, Sbragaglia2007, Leclaire2011, Fei2018e}.
	This treatment means that the interaction force is discretized with the 6th-order isotropic gradient~\citep{Shan2006, Leclaire2011}.
	Note that the isotropic gradient operator was first discussed in Shan~\citep{Shan2006}, which also gave the exact coefficients used here.
}

To incorporate a non-ideal equation of state (EOS), $p_{\mathrm{EOS}}$, into the pseudopotential model, the following potential form should be chosen as\footnote{Note that one can choose the potential form of $\displaystyle \psi(\boldsymbol{x}) = \sqrt{2(p_\mathrm{EOS}-\rho c_s^2)/\mathscr{G} c^2}$. 
	In this case, the pressure tensor in the recovered macroscopic equation will be slightly modified.}~\citep{He2002}
\begin{equation}
\psi(\boldsymbol{x}) = \sqrt{\frac{2(p_\mathrm{EOS}-\rho c_s^2)}{\mathscr{G} c_s^2}},
\label{eq:potential}
\end{equation}	
where $\mathscr{G}$ is only required to guarantee the term inside the square root to be positive~\citep{Yuan2006,Chen2014}.

Although the potential form of Eq.~(\ref{eq:potential}) is consistent with the EOS, the problem of thermodynamic inconsistency still persists~\citep{Li2016}.
To approximately restore the thermodynamic consistency, we here derive the corrected forcing term $\ket{R'_i}$ according to the idea of Li {\it et al.}~\citep{Li2012,Li2013b}.
To do this, the macroscopic velocity $\boldsymbol{u}$ in Eq.~(\ref{eq:forcingTerm}) is replaced by the following modified velocity:
\begin{equation}
\boldsymbol{u}' = \left\{ \begin{array}{ll}
\displaystyle \boldsymbol{u}+  \frac{\sigma \boldsymbol{F}_m}{(\omega^{-1}-1/2)\delta_t \psi^2} & (i=4,\,5), \\
\displaystyle \boldsymbol{u}+  \frac{\sigma \boldsymbol{F}_m}{(1-1/2)\delta_t \psi^2} & (\mathrm{otherwise}),
\end{array} \right.
\label{eq:modVel}
\end{equation}
where $\sigma$ is a constant to tune thermodynamic consistency.
By substituting Eq.~(\ref{eq:modVel}) into (\ref{eq:forcingTerm}) and transforming into central-moment space ({\it e.g.}, calculating $\bm{T} \ket{F_i(\boldsymbol{u}')}$), we can obtain the corrected forcing term.
Actually, we have found that $R_{3\dots8}$ are non-zero,
but for simplicity and consistency with the derived macroscopic equations (see Appendix \ref{sec:Analysis} for the analysis), we can modify only $R_3'$ and $R_8'$ as follows:
\begin{equation}
R'_3 = 4 \alpha |\boldsymbol{F}_m|^2,~ R'_8 = \frac{4}{3}\alpha|\boldsymbol{F}_m|^2,
\label{eq:force2}
\end{equation}
where $\alpha = \sigma /(\psi^2 \delta_t)$ and $|\boldsymbol{F}_m|^2 = F_{mx}^2+F_{my}^2$.
The other elements are kept to be the original ones ({\it e.g.}, $R_i\rq{}=R_i~\mathrm{for}~i=0\dots2~\mathrm{and}~i=4\dots7$).
Compared with the original forcing term in Eq.~(\ref{eq:force1}), the terms for $i=3\dots8$ have been modified.
When we choose $\sigma=0$, Eq.~(\ref{eq:force2}) reduces to the original forcing term of Eq.~(\ref{eq:force1}).
Due to this modification, we can rewrite the post-collision CMs as follows:
\begin{equation}
\begin{split}
k_0^{\star} = \rho,~ k_1^{\star} = \frac{1}{2}F_x,~ 	k_2^{\star} = \frac{1}{2}F_y,~ k_3^{\star} = \frac{2}{3}\rho \underline{+ 2 \alpha |\boldsymbol{F}_m|^2}, \\
k_4^{\star} = (1-\omega) k_4,~k_5^{\star} = (1-\omega) k_5, \\
k_6^{\star} = \frac{1}{6}F_y,~k_7^{\star} = \frac{1}{6}F_x,~ k_8^{\star} = \frac{1}{9}\rho \underline{+ \frac{2}{3}\alpha|\boldsymbol{F}_m|^2}.
\label{eq:postColl2}
\end{split}
\end{equation}
Comparing to Eq.~(\ref{eq:postColl}), the underlined terms have appeared in Eq.~(\ref{eq:postColl2}) to tune the thermodynamic consistency.

\subsection{Energy equation and non-ideal EOS}

	The energy equation employed in this paper, in terms of temperature $T$, is given by 
	\begin{equation}
	\frac{\partial T}{\partial t} + \boldsymbol{u}\cdot \nabla T 
	=  \frac{1}{\rho c_v} \nabla \cdot (\lambda \nabla T) - \frac{T}{\rho c_v} \left(\frac{\partial p_{\rm EOS}}{\partial T} \right)_\rho \nabla \cdot \boldsymbol{u},
	\label{eq:energy}
	\end{equation}
	where $\lambda$ is the thermal conductivity and $c_v$ is the specific heat at constant volume.
	One can derive the energy equation in the form of Eq.~(\ref{eq:energy}) with several approaches, {\it e.g.}, starting from the balance of internal energy or entropy.
	Details of the derivation can be found in the textbook~\citep{Slattery1972};
	discussion of the energy equation for phase-change LBMs can be found in Ref.~\citep{Li2018}.

To solve Eq.~(\ref{eq:energy}), we use the second-order Runge-Kutta scheme for time and	the third-order upwind scheme for the convection term\footnote{We stress that some {\it upwind}-type schemes should be adopted for the convection term. 
	Otherwise, the non-physical spurious oscillations will appear for convection-dominant cases.}.
The first-order derivative and the Laplacian are computed by the isotropic central difference~\citep{Shan2006,Sbragaglia2007} with the same stencils as the interparticle force in Eq.~(\ref{eq:interForce}).

As can be seen in Eqs.~(\ref{eq:interForce}) and (\ref{eq:energy}), the coupling between the pseudopotential model and the finite-difference scheme is established via a non-ideal EOS.
The realistic EOS widely used in the multiphase lattice Boltzmann simulations includes van der Waals, Carnahan-Starling, Peng-Robinson, etc.
We adopt the Peng-Robinson EOS in this simulations, which is given by~\citep{Yuan2006}
\begin{equation}
p_{\rm EOS} = \frac{\rho RT}{1-b\rho} - \frac{a\varphi(T)\rho^2}{1+2b\rho - b^2\rho^2}, 	\label{eq:eos}
\end{equation}
where $\displaystyle \varphi(T) = [1+(0.37464+1.54226\omega - 0.26992\omega^2) (1-\sqrt{T/T_c}) ]^2$ with the acentric factor $\omega$. 
The critical properties can be obtained as follows: $a=0.4572R^2T_c^2/p_c$ and $b = 0.0778RT_c/p_c$.
In this paper, we set $a=3/49$, $b=2/21$, $R=1$, and the acentric factor $\omega=0.344$ (for water).
Then, the critical density, temperature, and pressure can be calculated as  $\rho_c = 2.657$, $T_c = 0.1094$, and $p_c = 0.08936$, respectively.


\section{Numerical tests \label{sec:tests}}
In this section, we test the numerical properties of the above-outlined methodology.

\subsection{Four-rolls mill}
\begin{figure}
	\centering
	\includegraphics[width=8.2cm]{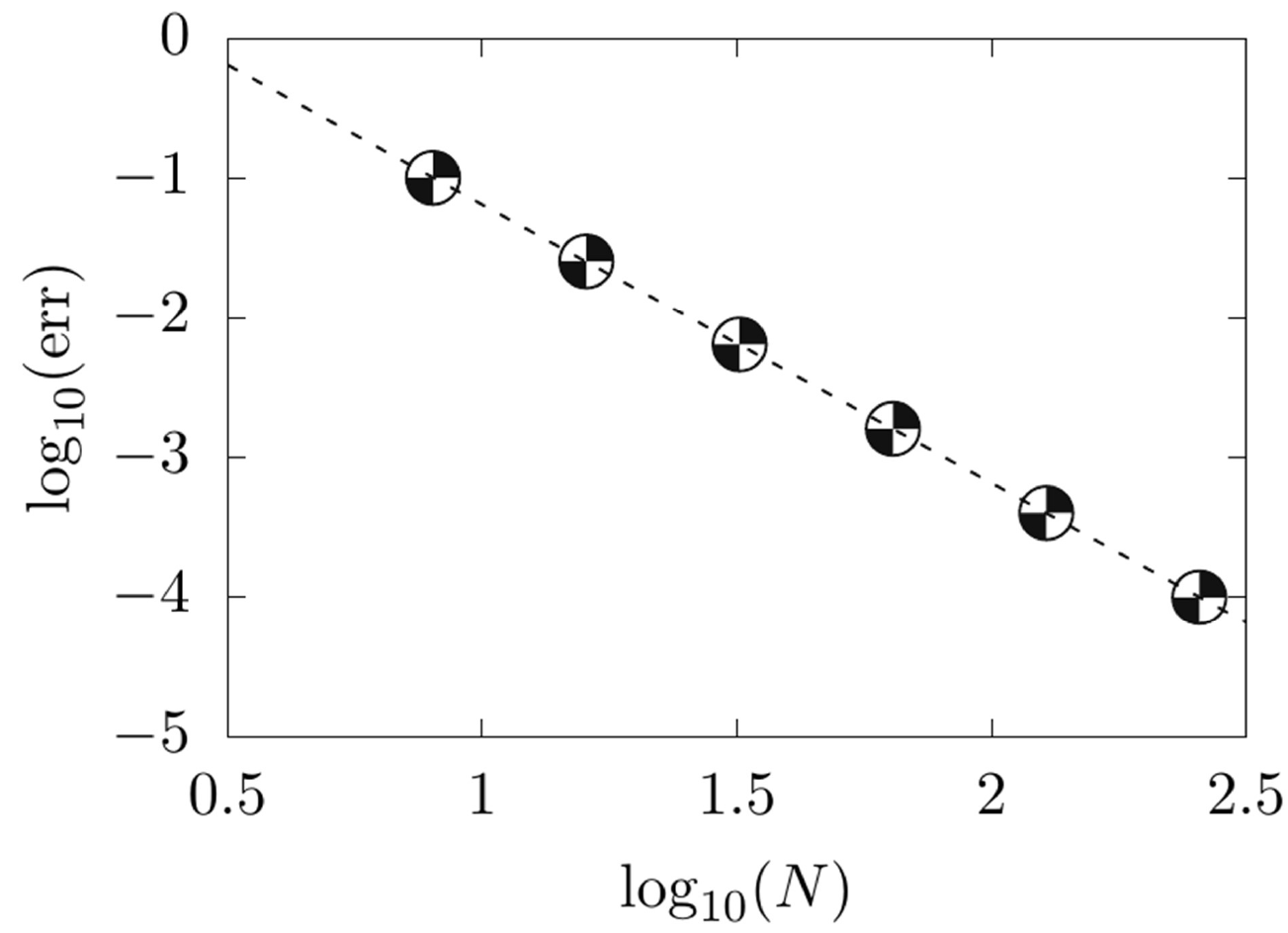}
	\caption{Four rolls mill: slope of the line fitting our results (triangles) indicates a convergence rate of 1.998.}
	\label{Figure1}
\end{figure}

Let us consider a square periodic box of size $N \times N$, where the fluid is initially at rest. 
Let us apply a constant force field, that is 
\begin{equation}
\boldsymbol{F}(\boldsymbol{x}) =  \phi \left[\sin(x) \sin(y),\cos(x) \cos(y)   \right].
\end{equation}
with $\phi = 2\nu u_0 \zeta^2$. Then, the pressure and velocity fields must converge to a steady state that reads as follows
\begin{equation}
\begin{split}
p (\boldsymbol{x}) &= p_0 \left[1-\frac{u_0^2}{4c_s^2} \left(  \cos \left(2\zeta x \right) -  \cos \left(2\zeta y \right)  \right)      \right],  \\
\boldsymbol{u}(\boldsymbol{x})  &= u_0 \left[\sin \left(\zeta x \right) \sin \left(\zeta y \right) , \,  \cos \left(\zeta x \right) \cos \left(\zeta y \right) \right], \label{tg_vel}
\end{split}
\end{equation}
where $u_0=10^{-3}$, $\zeta=2\pi/N$ and $p_0=\rho_0c_s^2$. 
This test is known as four-rolls mill, pioneered by \citet{Taylor1934}.
Simulations characterized by different grid sizes are carried out, \textit{i.e.} $N = \left[ 8,\, 16,\, 32,\, 64,\, 128 \right] $, at a Reynolds number $\displaystyle Re = u_0 N/\nu$ equal to 100. 
The performance of our scheme is elucidated by computing the relative discrepancy between analytical predictions and numerical findings. 
For this purpose, the vectors $\bm{r}_{\mathrm{an}}$ and $\bm{r}_{\mathrm{num}}$ are introduced, storing the values of the velocity field from Eq.~(\ref{tg_vel}) and those provided by our numerical experiments, respectively. Then, the relative error is computed as
\begin{equation}\label{error}
\mathrm{err} = \frac{\| \bm{r}_{\mathrm{an}}- \bm{r}_{\mathrm{num}}  \|}{ \| \bm{r}_{\mathrm{an}} \|}
\end{equation}
and is depicted in Fig.~\ref{Figure1} as a function of the grid dimension. 
An excellent convergence rate equal to 1.998 is found, which is totally consistent with the second-order nature of the LBE. 

\subsection{Double shear layer}
\label{DoubleShearLayer}


The double-shear layer benchmark is a very good test to evaluate the stability of the proposed numerical scheme~\cite{MINION_JCP_138_1997}. Given a periodic domain with $(x,y) \in [0,L]^2$, the flow develops due to the following initial conditions:
\begin{equation}
u_x(\bm{x},t=0) = \left\{
\begin{array}{ll}
u_0\, \mathrm{tanh}\left[\kappa \left(\frac{y}{L}- \frac{1}{4} \right) \right], \qquad \frac{y}{L} \leq \frac{1}{2},\\
\\
u_0\, \mathrm{tanh}\left[\kappa \left(\frac{3}{4} -\frac{y}{L} \right) \right], \qquad \frac{y}{L} > \frac{1}{2},
\end{array}
\right.
\end{equation}
and $\mathrm{Ma}=u_0/c_s=0.57$
\begin{equation}
u_y(\bm{x},t=0) = u_0 \delta \sin \left[2 \pi \left( \frac{x}{L}+\frac{1}{4} \right) \right],
\end{equation}
where $\kappa=80$ and $\delta=0.05$, representing two longitudinal shear layers and a superimposed transverse perturbation. The Reynolds number is $Re = u_0 L / \nu=3 \times 10^4$, with $L=512$. Three different values of the Mach number $Ma=u_0/c_s$ are considered: 0.1, 0.3 and 0.6.\\

\begin{figure}[htbp]
	\centering
	\includegraphics[width=8.2cm]{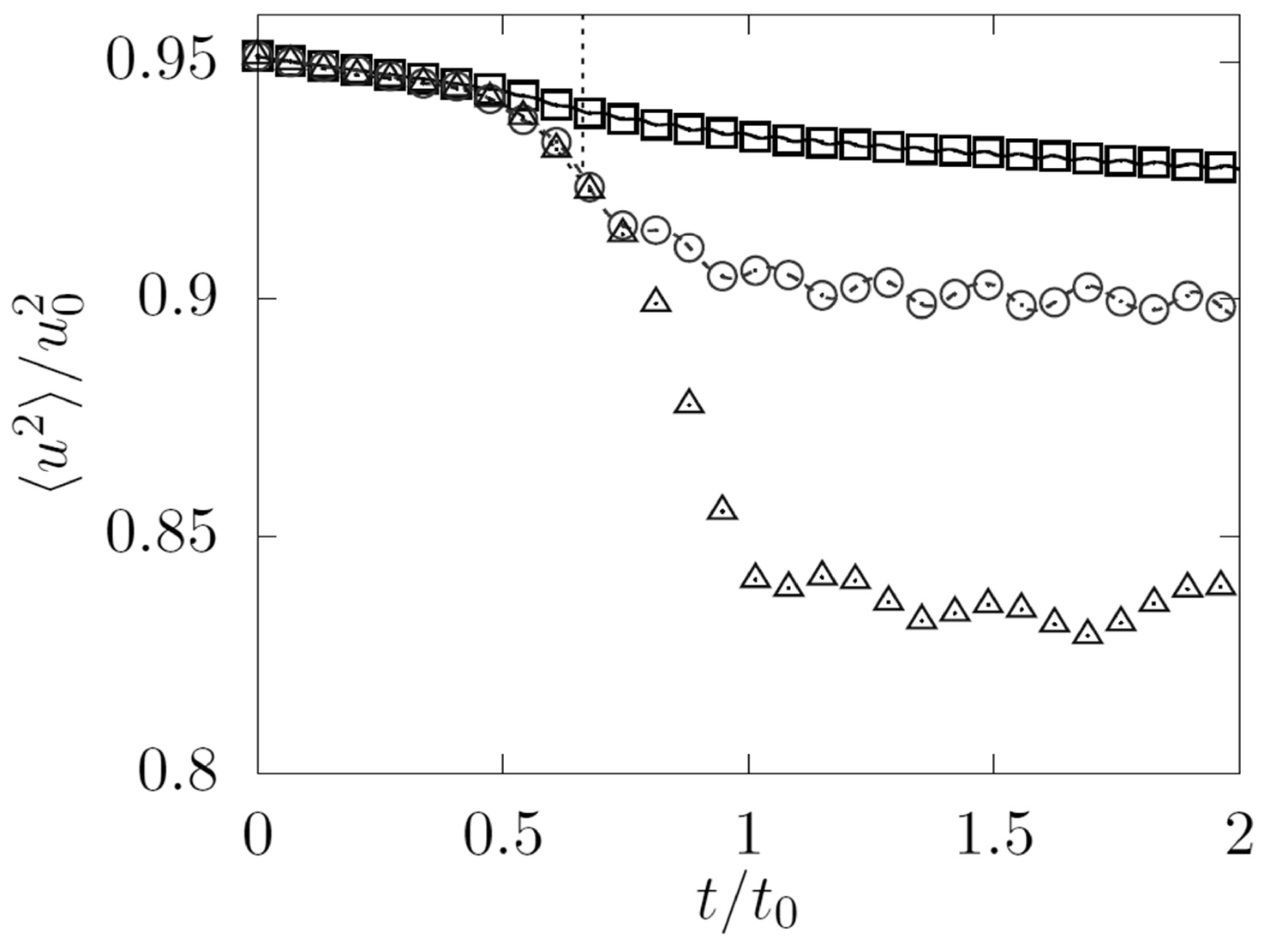}
	\caption{{Double shear layer: evolution of the normalized kinetic energy computed by the present scheme (lines) and  the BGK LBM (symbols) for different values of the Mach number: 0.1 (solid line and squares), 0.3 (dashed line and circles) and 0.6 (dotted line and triangles).}}
	\label{DSL}
\end{figure}

In Fig.~\ref{DSL}, the time history of the kinetic energy (normalized by its initial value) is reported by adopting the present scheme and the BGK LBM accounting for the full set of Hermite polynomials. As shown in the graph, results achieved by the present scheme are in excellent agreement with those obtained by the BGK LBM for the two lowest values of $\mathrm{Ma}$. Interestingly, the BGK LBM undergoes a sudden blow-up at $t/t_0 \sim 0.66$ for the case $Ma=0.6$, where $t_0 = L/u_0$, while our approach can simulate the whole desired time range, thus exhibiting remarkably higher stability.

\subsection{Poiseuille flow}
A simple yet effective test is represented by the Poiseuille flow. 
Let us consider a fluid initially at rest with density set to $\rho \left(\boldsymbol{x},0   \right)=\rho_0=1$ everywhere. 
A laminar flow develops in the $x$ direction within a domain of dimensions $L_x \times L_y$, that are respectively aligned to the $x,\,y$ Cartesian reference axis. No-slip boundary condition is enforced at two $y$-normal planes placed at the top and bottom sections of the domain, while the domain is periodic at the other two sides. By applying a constant uniform horizontal rightward force $F_x = 8 \nu U_0/L_y^2$, the flow field must converge to the analytical predictions
\begin{equation}
u_x(y) = -\frac{4U_0 y}{L_y}  \left( 1-\frac{y}{L_y} \right).
\end{equation}
The peak velocity of the imposed velocity profile is set to $u_0=0.001$ 	and the Reynolds number is $Re = u_0 L_y/\nu=100$. The value of $L_y$ varies in order to achieve different grid resolutions, \textit{i.e.} $L_y \in \left[5:1025 \right]$. Here, the simulations stop when the steady state is reached. In order to evaluate the accuracy and convergence properties of our algorithm, we define two vectors, $\bm{r}_{\mathrm{an}}$ and $\bm{r}_{\mathrm{num}}$ collect the analytical and numerical solutions, respectively.
\begin{figure}
	\centering
	\subfigure[]{\includegraphics[width=0.45\textwidth]{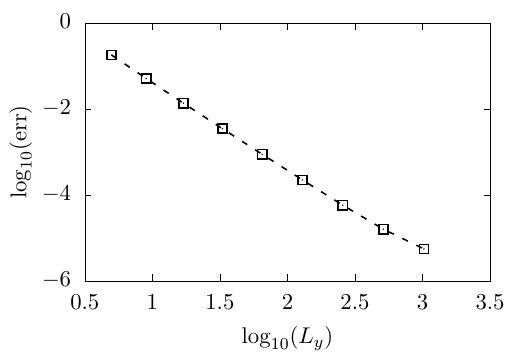}\label{PoiseuilleConvergence}}
	\subfigure[]{\includegraphics[width=0.45\textwidth]{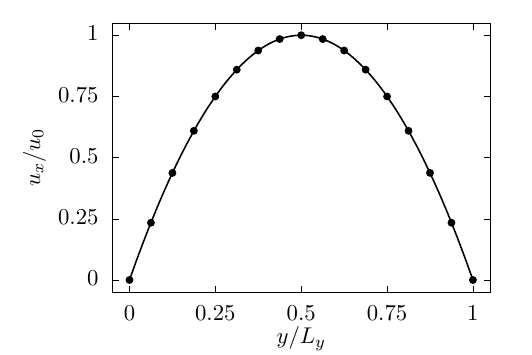}\label{PoiseuilleProfile}}
	\caption{Poiseuille flow: (a) convergence analysis showing an optimal convergence rate equal to 2.04 and (b) profile of the horizontal component of the velocity, $u_x$, normalized with respect to $u_0$ from the analytical predictions (continuous line) and our numerical scheme (circles).}
\end{figure}
In Fig.~\ref{PoiseuilleConvergence}, results obtained by a convergence analysis are reported. It is possible to appreciate that our algorithm shows an optimal convergence rate (\textit{i.e.} the slope of the line fitting the values) equal to 2.04, that is consistent with the second-order accuracy of the LBM. 
For $L_y = 1015$, the profile of the horizontal component of the velocity, $u_x/u_0$, from our numerical simulation is reported in figure \ref{PoiseuilleProfile}, together with the analytical predictions. It is possible to appreciate that the two solutions are overlapped.

\subsection{Static bubble}
To check the thermodynamic consistency of the present LBM scheme, static bubble tests are conducted.
In the computational domain discretized into $200 \times 200 $, 
periodic boundary conditions are set everywhere.
Following Li {\it et al.}~\citep{Li2013b}, we set the initial bubble radius to $R=50 $.
The analytical solution is obtained via the Maxwell equal area rule, {\it i.e.}, by numerical integration of the EOS.

\begin{figure}
	\begin{center}
		\includegraphics[width=7cm]{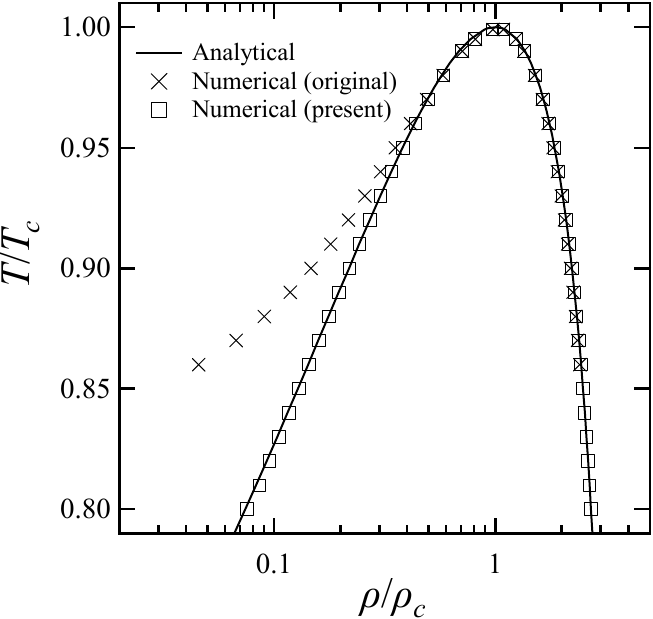}
		\caption	{Simulation of two-dimensional circular bubble with $\nu=0.1$. 
			Line: Maxwell's equal rule area. Symbol: lattice Boltzmann simulation with $\sigma = 0$ and $\sigma = 0.38$. 
			The case of $\sigma = 0$ corresponds to the original consistent scheme~\citep{Fei2017}.
			\label{fig:Thermodynamic}}
	\end{center}
\end{figure}

Fig.~\ref{fig:Thermodynamic} shows the comparison of simulation results with analytical solution of Maxwell construction.
As reference, the case with the original forcing scheme~\citep{Fei2017} is shown in the same figure.
From the Fig.~\ref{fig:Thermodynamic}, we can see that the present model well describes the analytical solution of Maxwell construction by setting $\sigma=0.38$, whereas the case with the original forcing scheme ($\sigma=0$) does not agree with the analytical solution.
Based on such observations, we confirmed that this model can approximately perform thermodynamic-consistent phase-change simulations.

\begin{figure}
	\begin{center}
		\includegraphics[width=7cm]{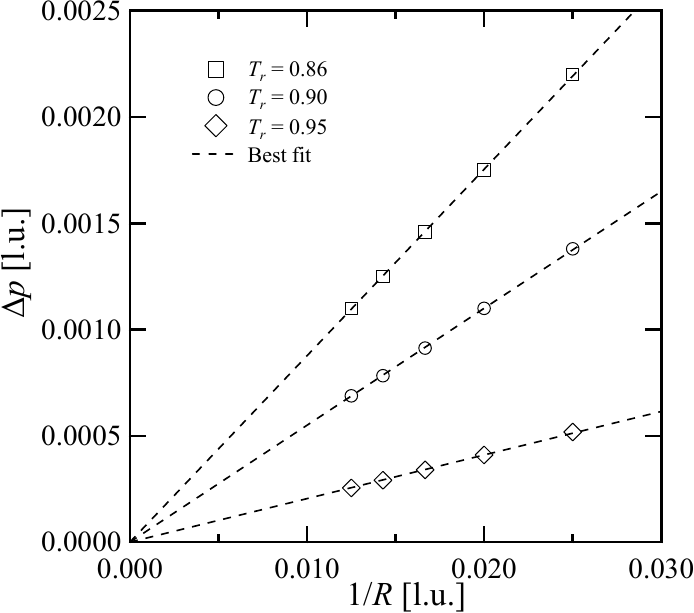}
		\caption{Laplace's law for several reduced temperatures: $T_r=0.86$, $T_r=0.90$, and $T_r=0.95$.
			\label{fig:Laplace}}
	\end{center}
\end{figure}

To validate the surface tension obtained by the present LBM scheme,  additional static-bubble tests are conducted.
We use the same computational setup as the last tests, and the parameter $\sigma$ is set to 0.38. 
A stationary bubble with various radius $R$ is initially placed in the domain.
From the Laplace law for two-dimensional case, the pressure difference between the inside and the outside of a bubble can be given by 
\begin{equation}
\Delta p = \frac{\gamma}{R},
\end{equation}
where $\gamma$ and $R$ are the surface tension and the bubble radius, respectively.
At the equilibrium state, we measure the pressure difference for three reduced temperature values, $T_r=0.86$, $0.90$, and $0.95$.
Fig.~\ref{fig:Laplace} shows the results.
As can be seen, our results well fit the Laplace law, that is, 
the pressure difference is proportional to $1/R$.

\subsection{Stefan problem}

\begin{figure}
	\begin{center}
		\includegraphics[width=8cm]{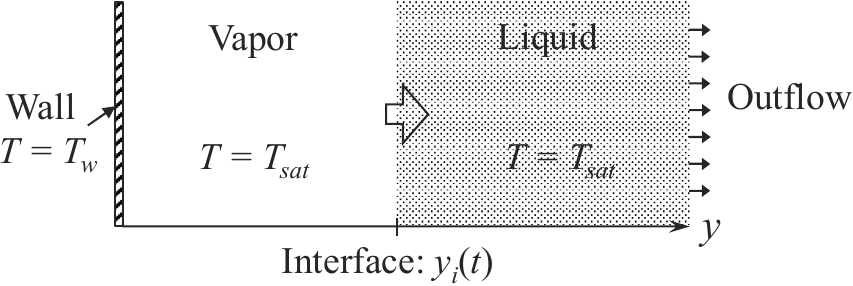}
		\caption{{Schematic representation of the Stefan problem.}
			\label{fig:schematic_stefan}}
	\end{center}
\end{figure}

	The last numerical test is the Stefan problem.
	This  benchmark problem has been solved by various numerical methods~\citep{Welch2000,Safari2013} to verify the time-dependent mass transport through the liquid-vapor interface.
	Schematic representation of the Stefan problem is shown in Fig.~\ref{fig:schematic_stefan}.
	When the wall temperature $T_w$ is higher than the saturation temperature $T_{sat}$, the interface moves along the $y$-axis due to evaporation of the liquid phase.
	The analytical solution for the interface location $y_i$ can be given by~\citep{Welch2000}
	\begin{equation}\label{eq:int_location}
		y_i(t) = 2\beta \sqrt{ \alpha_v t},
	\end{equation}
	where $\alpha_v$ is the thermal diffusivity of the vapor phase; $\beta$ is a solution to the following transcendental equation:
	\begin{equation}\label{eq:transcendental}
		\beta \exp({\beta^2}) \mathrm{erf}({\beta}) = \frac{c_{p,v} \Delta T}{h_{fg} \sqrt{\pi}},
	\end{equation}
	where $\Delta T (= T_w - T_{sat})$ is the superheat degree.
	Details of the derivation can be found in Ref.~\citep{Welch2000}.
	
	The computational domain is discretized into $W\times H= 5\times 100$, which represents a pseudo one-dimensional system.
	At the wall boundary, the bounce-back condition and the constant temperature $T=T_{w}$ are imposed.
	The outflow boundary condition is identical to the one in Sec.~\ref{sec:setup_boiling}.
	We started the computation with the initial interface location at $y_i(t=0)=5$ in lattice units.
	Here, we set $\alpha_v = 0.1$  in lattice units. 
	Calculation of the specific latent heat $h_{fg}$ is the same as the method described in Sec.~\ref{sec:setup_boiling}.
	
\begin{figure*}
	\begin{center}
		\includegraphics[width=12cm]{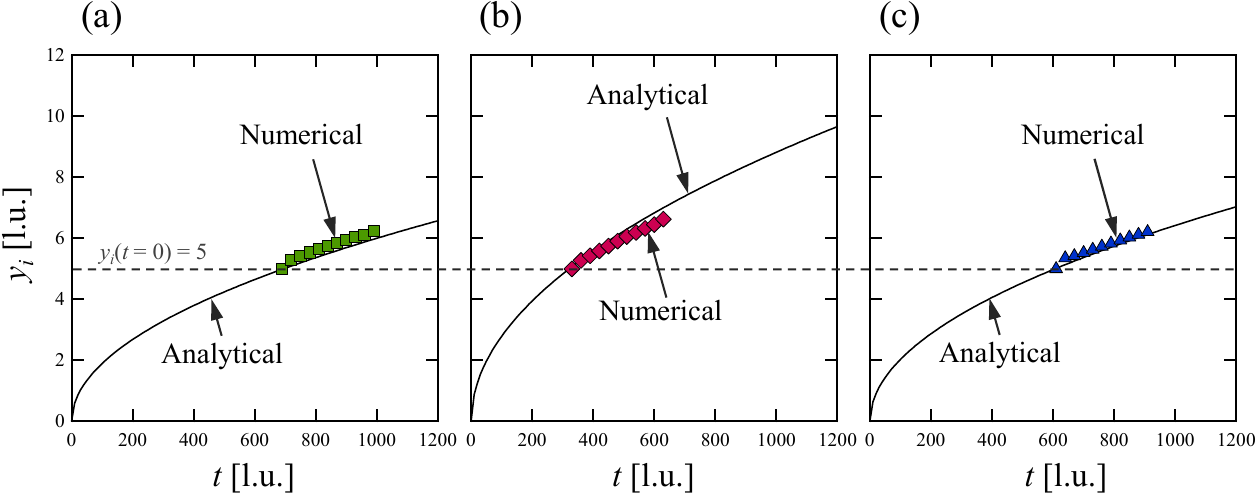}
		\caption{
			{Comparison of the present LBM results with the analytical solutions to the Stefan problem under several conditions: 
			(a) $T_r = 0.86$, $\Delta T = 0.010$; 
			(b) $T_r = 0.86$, $\Delta T = 0.023$;  
			(c) $T_r = 0.90$, $\Delta T = 0.010$. 
			Solid lines represent the analytical solution [Eq.~(\ref{eq:int_location})].}
			\label{fig:Stefan}}
	\end{center}
\end{figure*}

	Fig.~\ref{fig:Stefan} shows the comparison of numerical results with the analytical solution for different reduced temperature and superheat.
	We can see that for all conditions, the numerical solution is proportional to $\sqrt{t}$, which is the functional feature of the  analytical solution in Eq.~(\ref{eq:int_location}).
	The proportional constant $\beta$ obtained from the simulation is also in good agreement with the analytical solution.
	From these results, it can be confirmed that the time-dependent mass transport due to phase change can be calculated using the present LBM scheme.

\section{Simulations of forced-convection boiling \label{sec:simulation}}
In this section, we aim at simulating the melt particle settling in a coolant as described in Fig.~\ref{fig:fci}(b).
{
	Findings are shown both in lattice and physical units.
}

\subsection{Setup and initial parameters\label{sec:setup_boiling}}
\begin{figure}
	\begin{center}
		\includegraphics[width=5cm]{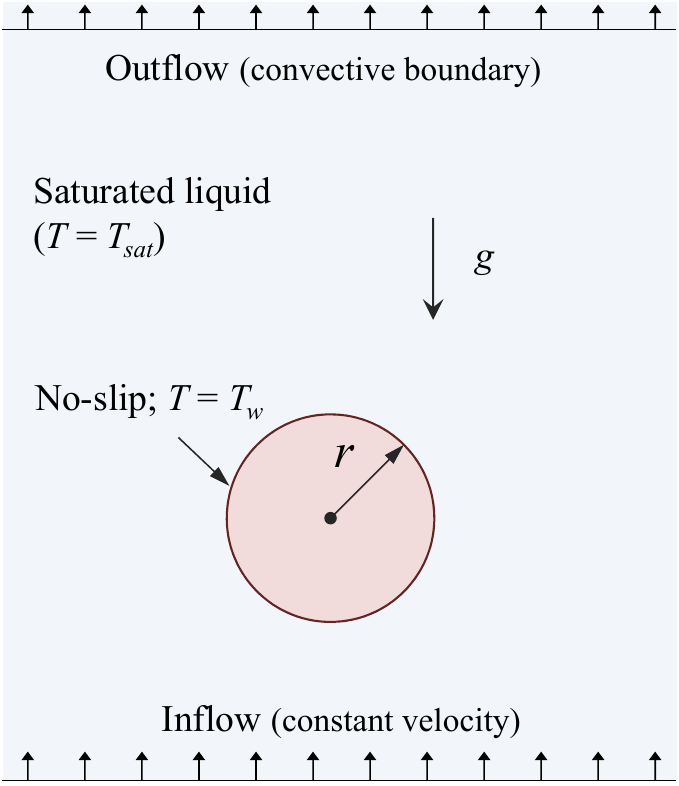}
		\caption[Boundary conditions for the boiling simulations.]
		{Boundary conditions for the boiling simulations. 
			\label{fig:boundary}}
	\end{center}
\end{figure}

The computational setup of the present simulations is described in Fig.~\ref{fig:boundary}. 
A constant velocity $u_l=0.05$ is prescribed at the inlet section~\citep{Zou1997}, while convective outlet is imposed at the outflow boundary~\citep{Lou2013}.
A cylindrical body is modeled by the interpolated bounce-back scheme~\citep{bouzidi2001momentum, DeRosis2014}, which is a similar approach to Hatani {\it et al.}'s condensation simulation~\citep{Hatani2015}.
In the cylindrical region with a diameter $D=30$, we set superheat temperature $T_w$, which is higher than the saturated temperature $T_{sat}$.
{
	In this research, we adopted the constant wall-temperature condition. 
	It should be noted that the boiling curves under the constant {\it wall-temperature} and constant {\it heat-flux} conditions would be identical in nucleate boiling and film boiling regimes, while they would differ only in the transition boiling regime~\citep{Zhang2017}.
}
The computational domain, filled with saturated liquid, is discretized into $W\times H= 20D\times 45D=600 \times 1350$.
The center of the cylinder is located at ($0.5W,0.2H$).
The initial liquid velocity is set to be $T_{sat}$ and $u_l$.
For the initial $3000$ time steps, we did not solve the energy equation and forcing term.

The liquid Reynolds number is an important parameter to investigate the effect of forced convection.
In the following simulations, two cases are considered: $Re=30$ and $30000$. 
The corresponding kinematic viscosity are $0.05$ and $0.00005$, respectively.
{According to Bromley {\it et al.}~\citep{Bromley1953}, the flow regime can be regarded as forced-convection boiling when $u_l/\sqrt{gD} > 2$.
	To set $u_l/\sqrt{gD}=4$, the gravitational acceleration is chosen as $g = 5.21 \times 10^{-6}	$ in this paper.}

In this paper, the reduced temperature is set to $T_r=T_{sat}/T_c = 0.86$.
{
	Note that this saturation temperature, $T_{sat}=0.86 T_c$, is higher than $27/32T_c \sim 0.84T_c$, which is the homogeneous nucleation temperature based on the foam-limit theory of \citet{Spiegler1963}.}
In this case, the saturation temperature corresponds to 556 K in actual physical units and the system pressure to $\sim$7 MPa.
The liquid-vapor density ratio results in $\rho_l/\rho_v \sim 17$.
The physical properties required for the simulations (kinematic-viscosity ratio, Prandtl number, thermal conductivity ratio, specific heat, etc.) under the reduced temperature are taken from the steam table for the saturation condition~\citep{JSME}, which gives the Prandtl number for vapor phase equal to $Pr=1.5$.
We have assumed that specific heat ratio for each phase is unity, and we set to $c_{p,v}=11$.
The specific latent heat, $h_{fg}$, is determined from the EOS with the same procedure in \citet{Gong2013}. 
Our parameter choice of $a$ and $b$ in Eq.~(\ref{eq:eos}) leads to $h_{fg}=0.572$. 
The wall superheat $\Delta T$ is varied from $0.005$ to $0.065$.

{
In our boiling simulations, the mesh size  $\Delta x$ in real units can be calculated as 0.031 mm. 
The time step size $\Delta t$ can be calculated as 0.387 ms and 0.000387 ms for $Re$ = $30$ and $30000$, respectively.
The conversion of lattice to physical units for several quantities, including the mesh size $\Delta x$ and time step size $\Delta t$, is summarized in Appendix~\ref{sec:Conversion}.}

\subsection{Boiling regimes, heat flux, and boiling curve}

\begin{figure*}
	\begin{center}
		\includegraphics[width=0.8\linewidth]{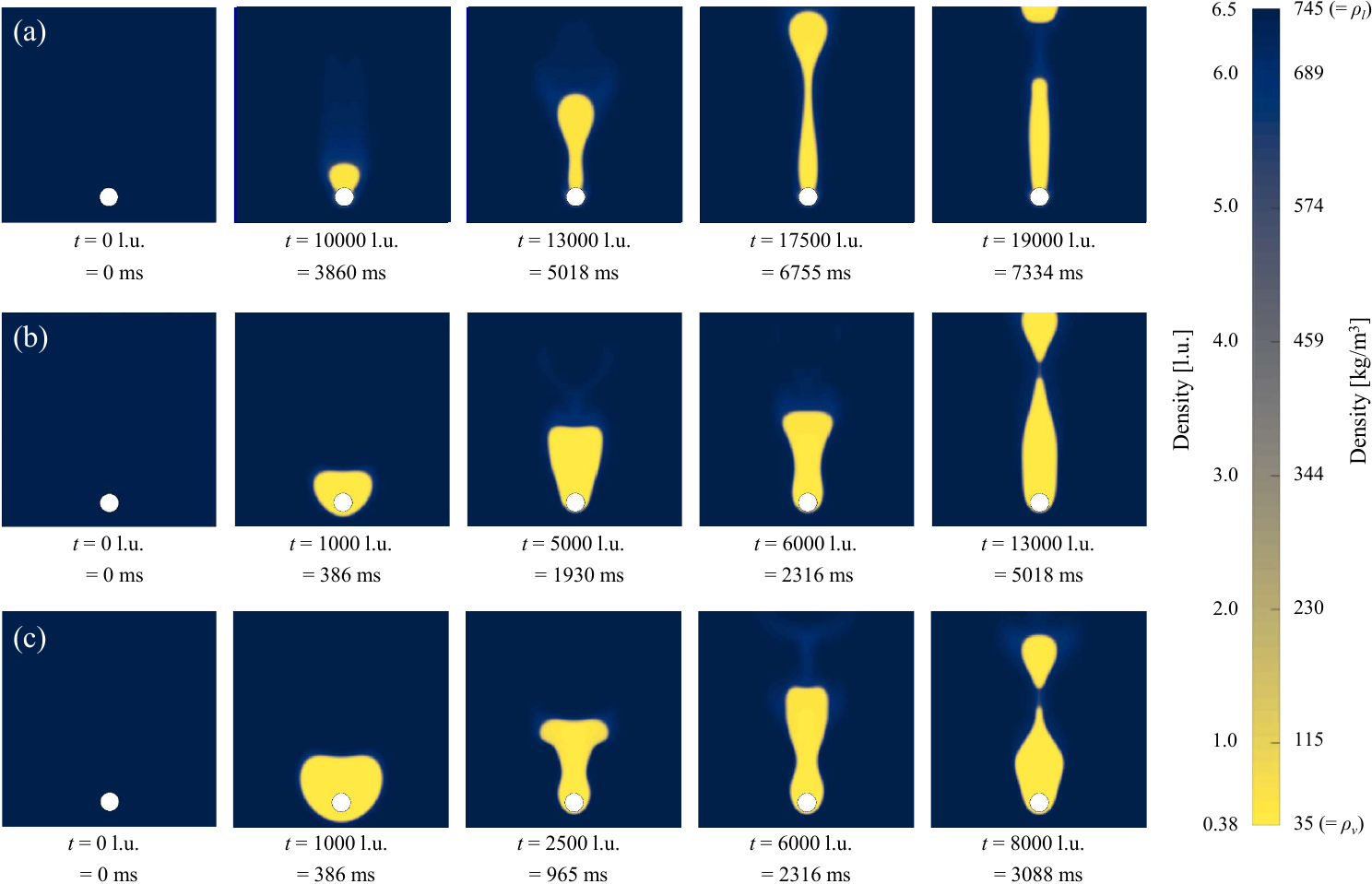}
		\caption	{ Snapshots of {density distribution during} convection boiling on a cylindrical body at $Re = 30$  and $Pr = 1.5$: 
				{(a) $\Delta T = 0.010$ l.u. ($=59$ K), 
				(b) $\Delta T = 0.023$ l.u. ($=137$ K), and 
				(c) $\Delta T = 0.060$ l.u. ($=356$ K).}
			\label{fig:snapBoilRe30}}
	\end{center}
\end{figure*}

\begin{figure*}
	\begin{center}
		\includegraphics[width=0.8\linewidth]{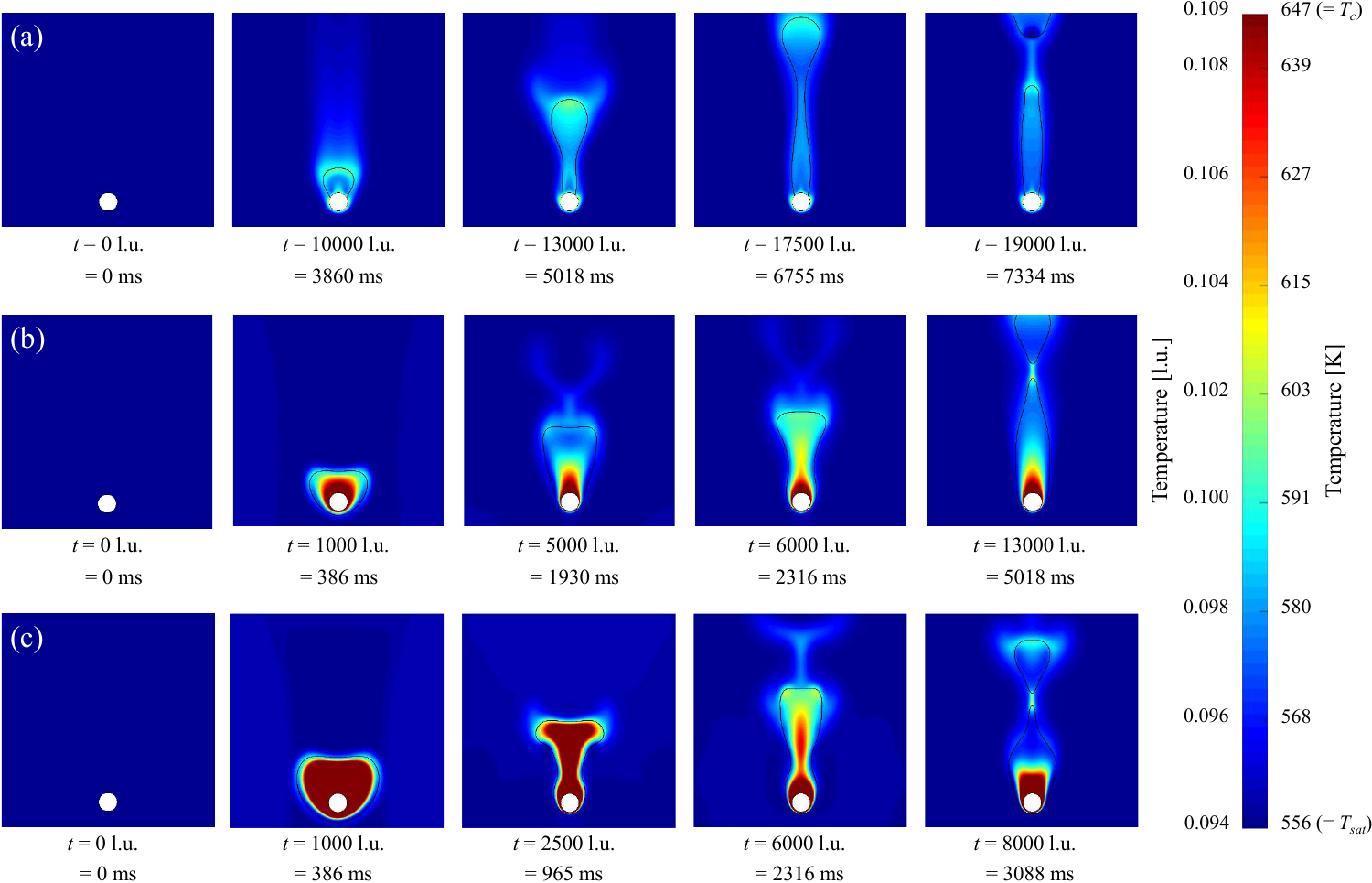}
		\caption	{ {Snapshots of temperature distribution during convection boiling on a cylindrical body at $Re = 30$  and $Pr = 1.5$: 
			(a) $\Delta T = 0.010$ l.u. ($=59$ K), 
				(b) $\Delta T = 0.023$ l.u. ($=137$ K), and 
				(c) $\Delta T = 0.060$ l.u. ($=356$ K).
			Solid lines in the pictures represent the interface defined by $(\rho_l+\rho_v)/2$.}
			\label{fig:snapTempRe30}}
	\end{center}
\end{figure*}

Simulation results for $Re=30$ are shown in {Figs.~\ref{fig:snapBoilRe30} and \ref{fig:snapTempRe30}}.
Density {and temperature} distributions around a cylinder with different superheat degrees are cut out and magnified.
All the simulations started with being fully immersed by liquid phase. 
After a certain time elapses, nucleation naturally arises around the superheated cylindrical domain, confirming the capability of the present approach to capture this feature.
The appeared vapor phase grows up and finally breaks up in different forms, depending on the superheat degree.
For the low superheat case [Fig.~\ref{fig:snapBoilRe30}(a)], it seems that only the upper part of the cylinder is in contact with the vapor phase.
In contrast, the high superheat cases [Fig.~\ref{fig:snapBoilRe30}(c)] implies that most of the cylinder's surface is covered with vapor phase.
This may be recognized as film-boiling regime.
{From the temperature distributions for this regime [Fig.~\ref{fig:snapTempRe30}(c)], we can see that the temperature around the cylinder is very high; while, the temperature downstream of the cylinder is low.}

\begin{figure*}
	\centering
	\includegraphics[width=0.7\linewidth]{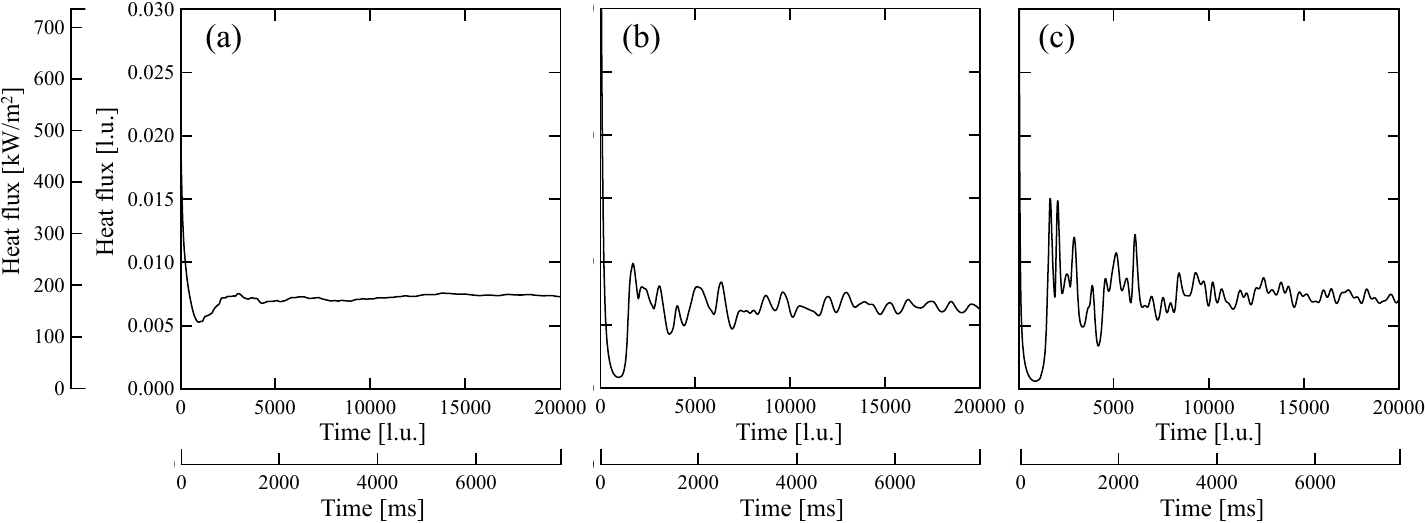}
	\caption	{ Time histories of heat flux on a cylindrical body  at $Re = 30$  and $Pr = 1.5$: 
			{(a) $\Delta T = 0.010$ l.u. ($=59$ K), 
			(b) $\Delta T = 0.023$ l.u. ($=137$ K), and 
			(c) $\Delta T = 0.060$ l.u. ($=356$ K).} 
		\label{fig:heatFlux1}}
\end{figure*}

The above observations of boiling regimes are qualitative.
In order to perform a more quantitative investigation, we calculate the average heat flux on the cylinder surface, that is
\begin{equation}
\bar{q}(t) = \frac{1}{W}\int_0^W{q(\bm{x})}\mathrm{d}s,
\end{equation}
where $W$ is the circumference length, and
the local heat flux is given by
\begin{equation}
q(\bm{x}) = -\lambda (\bm{x}) {\frac{\partial T}{\partial n}}|_{\mathrm{wall}},
\end{equation}
where $\lambda$ is the thermal conductivity and $n$ is the normal direction.
Fig.~\ref{fig:heatFlux1} represents the time history of the calculated  heat flux.
Simulation conditions are the same as Fig.~\ref{fig:snapBoilRe30}.
For low superheat case [Fig.~\ref{fig:heatFlux1}(a)], the heat flux data seems to converge to a certain value.
As the superheat is increased, the heat flux starts to intensely fluctuate.
This is due to intermittent contact of superheated wall and liquid.
When it is in contact with the liquid, higher heat flux is observed;
when it is in contact with the vapor, lower heat flux is observed.
We find that the intervals of this intermittent fluctuation shortened with the increase of superheat.

\begin{figure}
	\centering
	\includegraphics[width=1.0\linewidth]{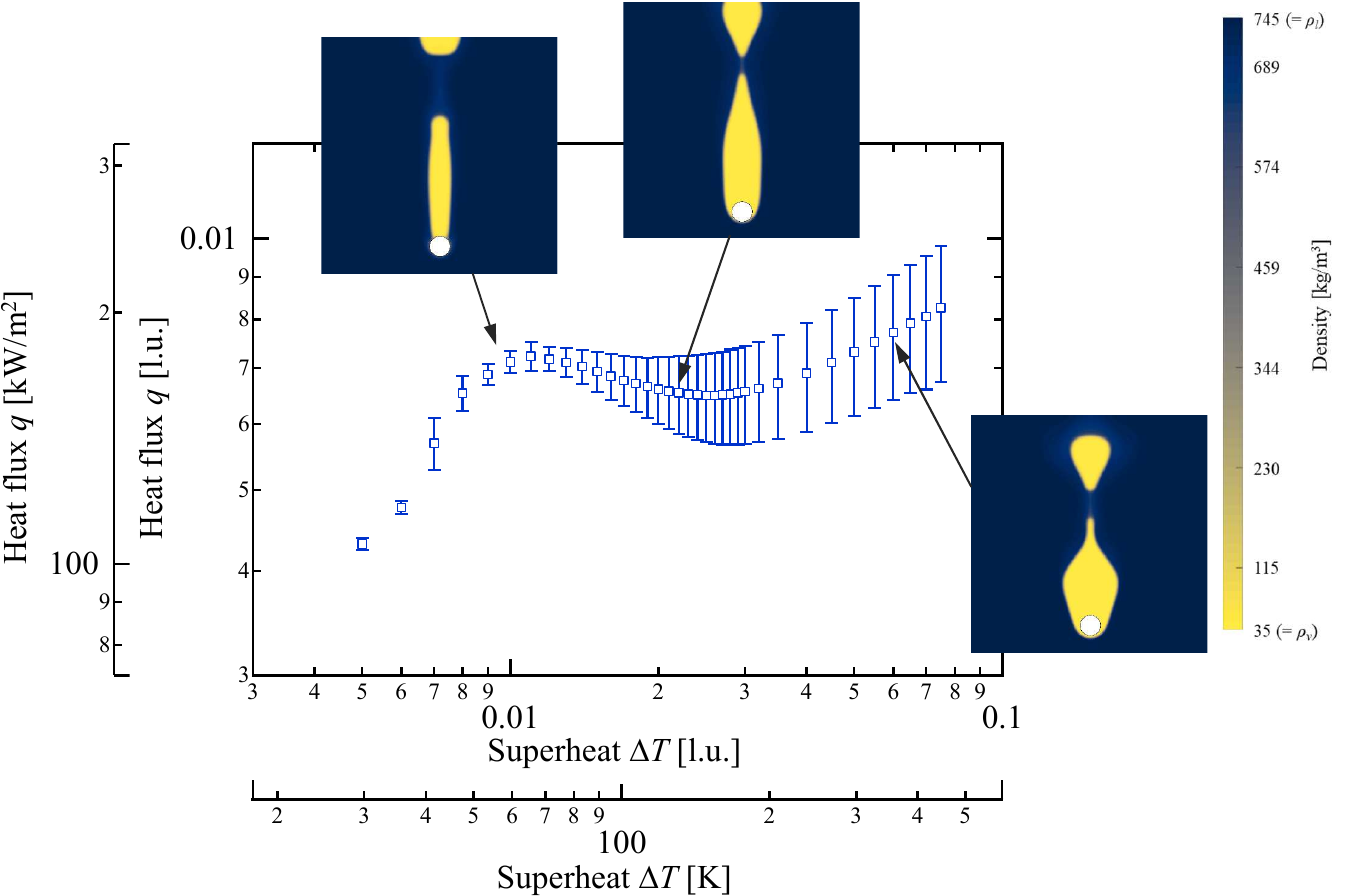}
	\caption{
		Boiling curve for $Re = 30$ and $Pr = 1.5$.
		{Snapshot of density distributions for $\Delta T = 0.010$ l.u. ($=59$ K), 
		$\Delta T = 0.023$ l.u. ($=137$ K), and 
		$\Delta T = 0.060$ l.u. ($=356$ K)} are included in the figure.
	}
	\label{fig:boilingcurvere30}
\end{figure}

By averaging the heat flux during the quasi-steady state, we can calculate the time-averaged heat flux.
When the heat flux is plotted against the superheat, the obtained curve describes the boiling curve, which is also known as the Nukiyama curve.
Fig.~\ref{fig:boilingcurvere30} shows the boiling curve for $Re = 30$. 
The error bar indicates the standard deviation of the time-averaged heat flux.
Typical snapshots at $\Delta T = 0.010$, $0.023$, and $0.060$ (in lattice units) are displayed in Fig.~\ref{fig:boilingcurvere30}.
The tendency of the heat flux against superheat degree agrees with that in Fig.~\ref{fig:typicalboilingcurve}.
At low superheat conditions, the heat flux monotonously increases and reaches the maximal point.
Although the present system is the convection boiling, the boiling regime in this low superheat region can be regarded as nucleate boiling. 
The maximal point would be the CHF.
After the CHF point, the heat flux gradually decreases and reaches the minimal point, MHF.
The boiling regime between the CHF and MHF would correspond to transition boiling.
Finally, the heat flux is found to increase with the superheat degree over the MHF.
This boiling regime would be the film boiling.
We stress here that the present simulation can deal with a series of boiling processes, {\it i.e.},  vapor bubbles nucleation, growth, and departure. 
Since all the boiling regimes can be reproduced, the results are  not limited to the film-boiling regime without any artificial input unlike Navier--Stokes-based modeling~\citep{Son1997,Juric1998,Welch2000,Yuan2008,Phan2018}.

\begin{figure}
	\centering
	\includegraphics[width=1.0\linewidth]{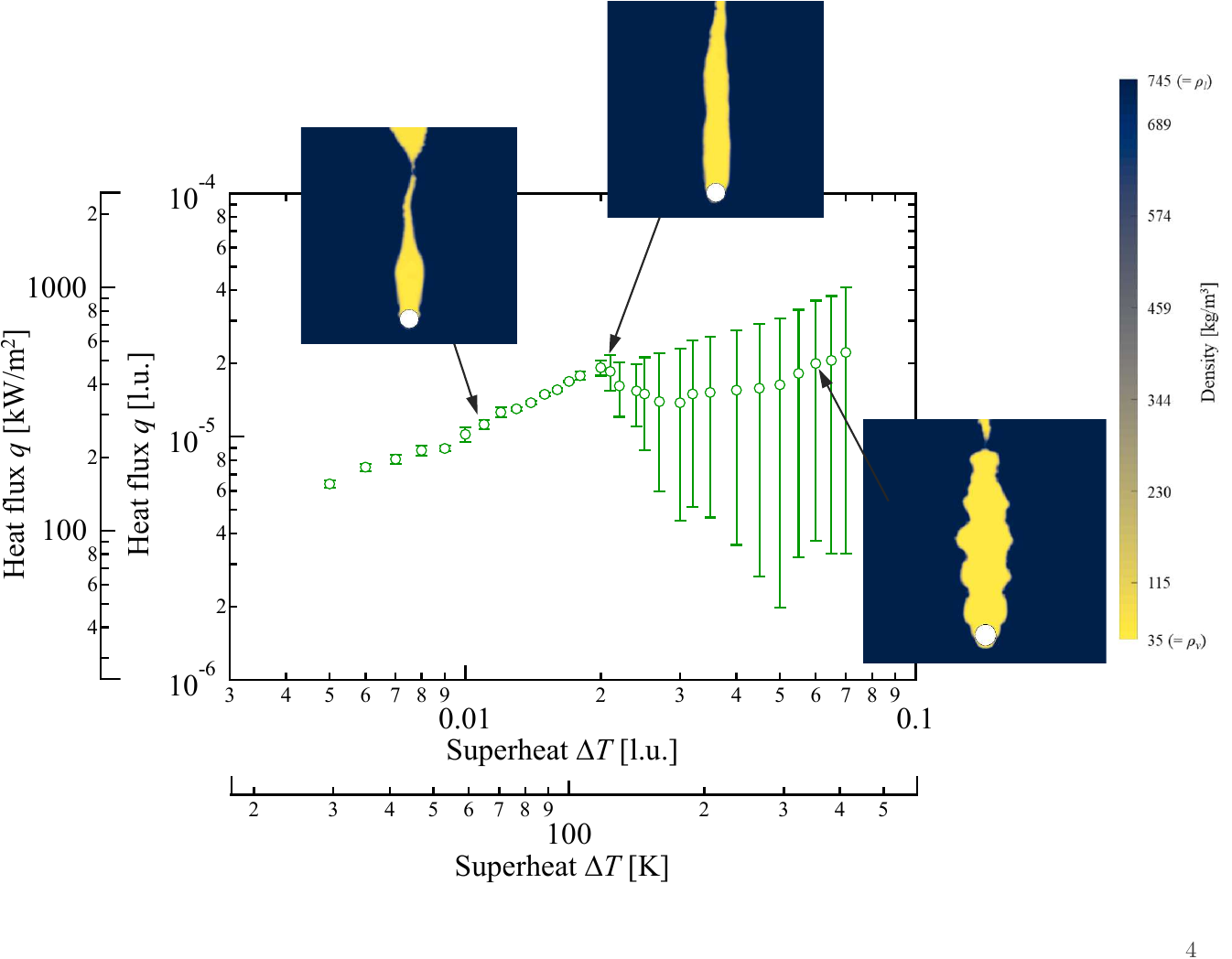}
	\caption{
		Boiling curve for $Re = 30000$ and $Pr = 1.5$.
		{Snapshot of density distributions for 
		$\Delta T = 0.012$ l.u. ($=71$ K),  
		$\Delta T = 0.021$  l.u. ($=125$ K), and 
		$\Delta T = 0.060$  l.u. ($=356$ K)} are included in the figure. 
		\label{fig:BoilingCurveRe30000}}
\end{figure}

Fig.~\ref{fig:BoilingCurveRe30000} shows the boiling curve for $Re = 30000$ with three snapshots of boiling behavior. 
The averaged heat-flux values indicate that there can be found the maximal (CHF) and minimal  (MHF) points in Fig.~\ref{fig:BoilingCurveRe30000}.
It should be noted that our CMs-based LBM approach enables to stably perform forced-convection boiling simulations up to very high $Re$ with the order of $O(10^4)$.
The error bars of heat flux beyond the CHF point are much larger than the case for $Re=30$ in Fig.~\ref{fig:boilingcurvere30}, 
since the inertia of the liquid phase increases.
As mentioned above, this is due to the intermittent contact with vapor and liquid.
This behavior promotes an unstable interfacial deformation as can be seen in the snapshot at $\Delta T =0.060$ within Fig.~\ref{fig:BoilingCurveRe30000}.
From the boiling curve characteristics, we can recognize that the film boiling begins when $\Delta T$ is around 0.03.
However, the present simulations indicate that the cylindrical body intermittently contacts with liquid.
Such behavior was also slightly seen even at $Re = 30$. 
In case of $Re = 30000$, the solid-liquid contact was promoted mainly because the inertial force of forced convection was larger than the force driven by vaporization around the cylinder.
Destabilization of the interface would also help the intermittent direct solid-liquid contacts.
Actually, experimental observations show that the local and intermittent solid-liquid contact occurs even in the film-boiling regime~\citep{Bradfield1966, Yao1978, Kikuchi1992}.
Such solid-liquid direct contacts were actually confirmed from the present simulations.
Our results support the experimental considerations of solid-liquid direct contact even in the film-boiling regime.

\begin{figure*}
	\begin{center}
		\includegraphics[width=0.6\textwidth]{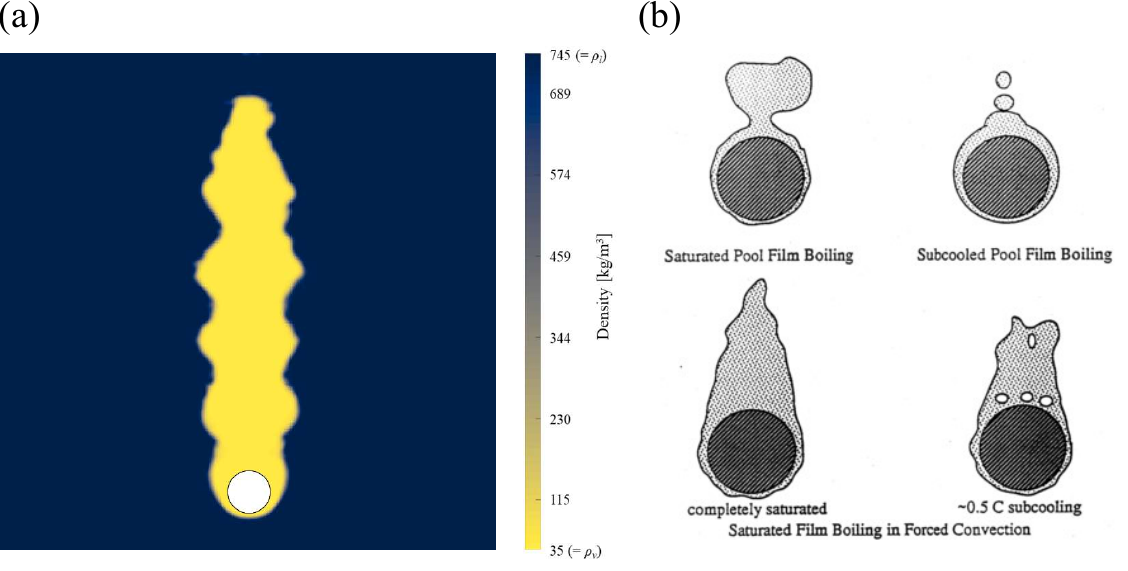}
		\caption{
			Comparison of simulation result and conceptual illustration of several film-boiling regimes reproduced from Liu and Theofanous~\citep{Liu1994}.
			(a) Snapshot of simulation result for $Re=30000$, (b) Conceptual illustration of film-boiling regime, including saturated pool film boiling, subcooled pool film boiling, completely saturated film boiling in forced convection, and slightly subcooled film boiling in forced convection.
			\label{fig:comparisonWithLiu}}
	\end{center}
\end{figure*}

We then compare the present simulation results with the conceptual illustration of several film-boiling regimes reported by Liu and Theofanous~\citep{Liu1994}.
They schematically illustrated characteristic film-boiling regimes, including saturated and subcooled pool film boiling and completely saturated and slightly subcooled film boiling in forced convection.
Fig.~\ref{fig:comparisonWithLiu} shows one of the typical simulation snapshots of film boiling for $Re=30000$ and Liu--Theofanous' conceptual illustration of film-boiling regimes.
From the simulation result [Fig.~\ref{fig:comparisonWithLiu}(a)], we can see that the vapor is stretched into downstream region due to the vaporization and forced convection.
Most of the cylindrical body is covered with vapor phase.
Globally, the present simulation seems to capture the characteristics of the type of ``completely saturated" film boiling in forced convection as in Fig.~\ref{fig:comparisonWithLiu}(b), although a slight difference in the interfacial shape still remains.	

\subsection{Comparison with film boiling heat transfer correlations}

Focusing on the heat-flux data of the film-boiling regime, one can compare the present simulation with the existing film boiling HTC correlations.
{
	We decided to use only the HTC correlations for the system of film boiling on a cylinder~\citep{Bromley1953,Epstein1980,Ito1981} to be consistent with our two-dimensional simulations.}
All the correlations considered here have been reduced for the saturation condition, i.e., the term related subcooling degree has been dropped.
The functional form of them are summarized in Appendix~\ref{sec:correlations}.

\begin{figure*}
	\centering
	\includegraphics[width=0.8\linewidth]{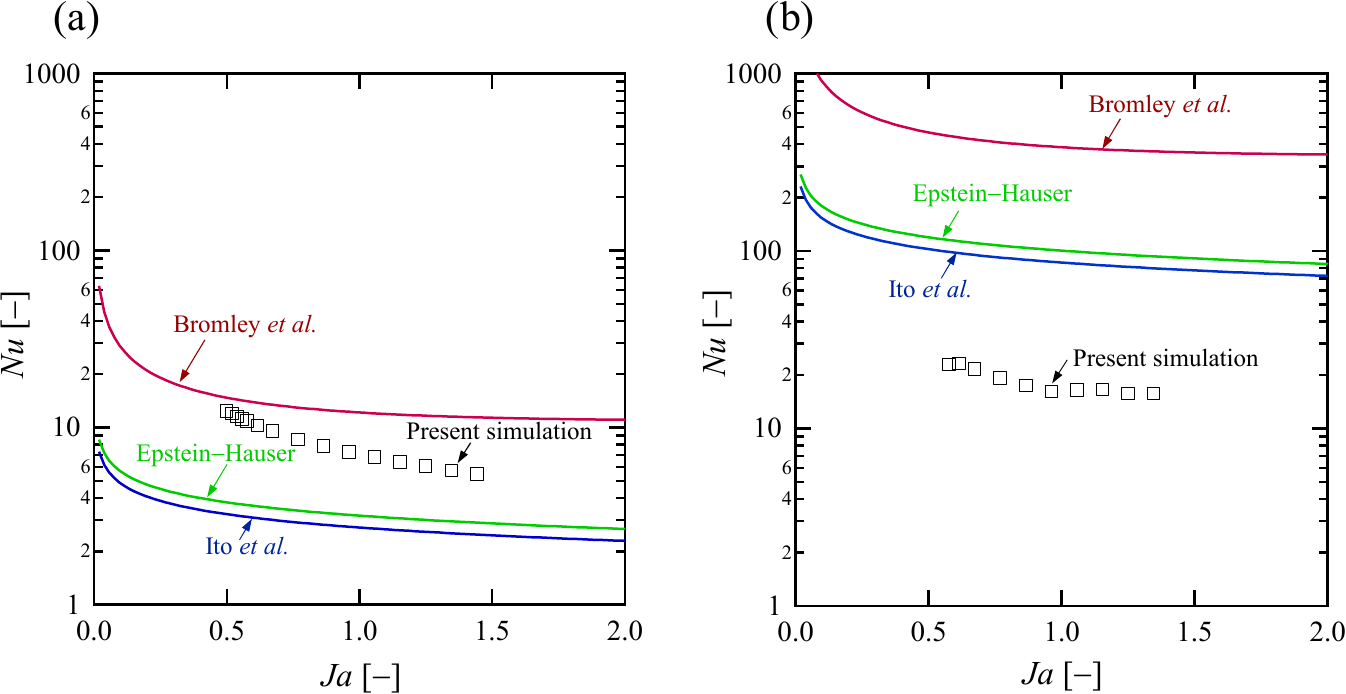}
	\caption
	{
		Comparison of the simulation results (open symbols) with the Mode 1~\citep{Bromley1953} and Mode 2~\citep{Epstein1980,Ito1981} correlations for cylinders.
		(a) $Re=30$ and (b) $Re=30000$.
		\label{fig:correlation_Re30_and_Re30000}}
\end{figure*}

The Nusselt number, ${Nu}$, calculated from the simulation results is shown in Fig.~\ref{fig:correlation_Re30_and_Re30000} for $Re=30$ and $Re=30000$.
For both low- and high-$Re$ data, ${Nu}$ decreases as ${Ja}$ increases.
In Fig.~\ref{fig:correlation_Re30_and_Re30000}, six kinds of correlations are also shown together with the simulation data.
{These can be classified as Mode 1~\citep{Bromley1953} and Mode 2~\citep{Epstein1980,Ito1981} correlations.
	On the ${Nu}$ and ${Ja}$ numbers, the Mode 1 correlations state that ${Nu} \propto {Ja}^{-1/2}$, while the Mode 2 correlations show ${Nu} \propto {Ja}^{-1/4}$.}
Let us first focus on the low-$Re$ case [Fig.~\ref{fig:correlation_Re30_and_Re30000}(a)].
{The $Nu$ obtained from the simulation appears in between the Bromley {\it et al.} and Epstein--Hauser correlations.
	The order of $Nu$ could be predicted by all correlation considered here, but the slope was slightly different, especially when the Jacob number was small.}
Then we turn to the high $Re$ case [Fig.~\ref{fig:correlation_Re30_and_Re30000}(b)]. 
{All the correlations overestimate the ${Nu}$ obtained from the present simulations, but the slope of the simulation data was close to the Epstein--Hauser and Ito {\it et al.} correlations (Mode 2 correlations).}
This means that adjusting the coefficient $C$ in the correlations would corroborate the simulation results.
The coefficient $C$ was actually used as a fitting parameter in past experiments.
Eventually, the present comparison of simulations and correlations implies that one should carefully choose the correlation depending on system's Reynolds number, and it would be difficult to identify a unique (optimal) solution for different flow conditions.

\begin{figure*}
	\centering
	\includegraphics[width=0.7\linewidth]{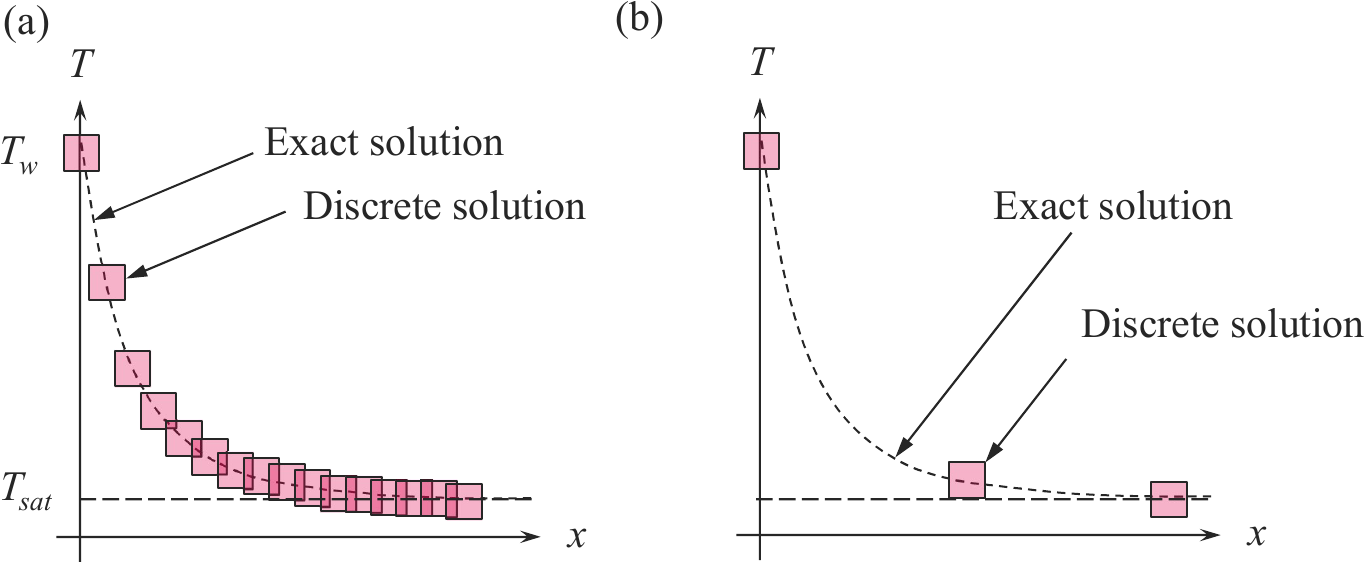}
	\caption{One-dimensional concept of the relation between exact and discrete solutions: (a) low Reynolds number (b) high Reynolds number. }
	\label{fig:boundarylayer}
\end{figure*}

Although our proposed numerical method enables the stable simulation at an extremely high Reynolds number ($Re = 30000$), the obtained heat flux is lower than the one given by all the afore-mentioned correlations.
One of the reasons should be found in the decrease of thermal boundary layer when the Reynolds number is high.
In our simulations, the Prandtl number $Pr$ is fixed at constant value regardless of the Reynolds numbers.
The expected situation is schematically illustrated in Fig.~\ref{fig:boundarylayer} as a one-dimensional relation between the exact and discrete solutions on lattice grids.
Let us assume the grid resolution is the same and the Prandtl number is fixed at a same value.
For low-$Re$ case [Fig.~\ref{fig:boundarylayer}(a)], the thermal boundary layer can be easily resolved with the discrete grid.
In contrast, the high $Re$ [Fig.~\ref{fig:boundarylayer}(b)] would have lead to very thin thermal boundary layer.
The fact makes it difficult to accurately capture the temperature gradient near the wall, and the evaluated value of temperature gradient will underestimate the exact one.
{
	To quantitatively support this argument, we performed simulations of bubble growth in an overheated liquid. 
	A description of the simulation details is available in Appendix~\ref{sec:bubbleGrowth}. 
	The results showed insufficient spatial resolution at low thermal diffusivity (corresponding to the high-$Re$ case: $Re$ = 30000 and $Pr$ = 1.5).}
{
	This bubble-growth simulation also suggests that the fluctuations of the heat flux over one order of magnitude in Fig.~\ref{fig:BoilingCurveRe30000} would not be correctly captured due to insufficient resolution of the thermal boundary layer.
}

\begin{figure*}
	\centering
	\includegraphics[width=0.55\linewidth]{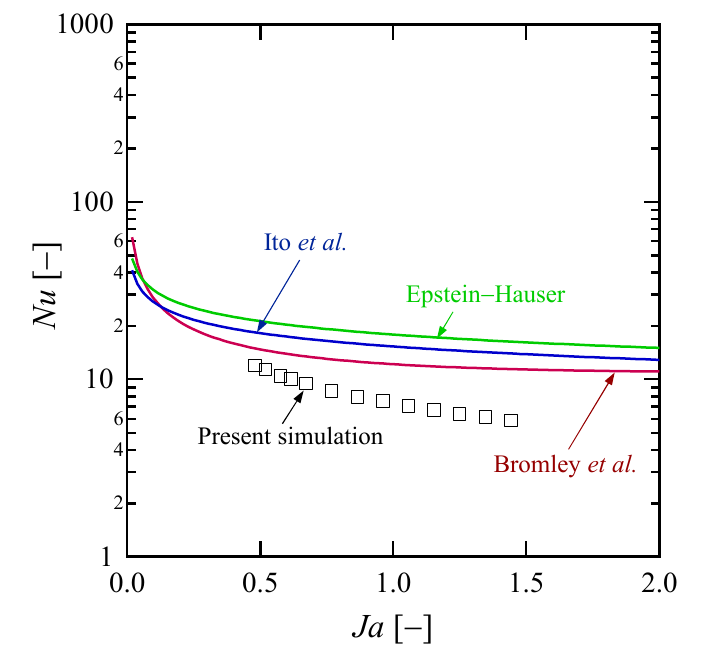}
	\caption{
		Comparison of the low-Prandtl-number simulation results (open symbols) with the Mode 1~\citep{Bromley1953} and Mode 2~\citep{Epstein1980,Ito1981} correlations for cylinders. 
	}
	\label{fig:correlationre30000lowpr}
\end{figure*}

To examine this assumption, we conducted additional simulations with extremely low Prandtl number ($Pr = 1.5/1000$), while keeping the high Reynolds number $Re = 30000$.
By setting the virtually low Prandtl number, we aim to artificially increase the thickness of thermal boundary layer.
Fig.~\ref{fig:correlationre30000lowpr} again plots the simulation results and correlations on the ${Nu}$-$Ja$ diagram.
{Compared with the previous simulations with high $Re$ [Fig.~\ref{fig:correlation_Re30_and_Re30000}(b)],
	the order of agreement is improved.
	The Mode 1 correlation is close to the simulations in terms of $Nu$ values.
}
Virtually lowering $Pr$ is considered to have improved the heat-flux calculation around the cylindrical wall.
All in all, the present examination proves the importance of the thickness of thermal boundary layer in heat flux calculation.
{As a result of improving the thermal boundary layer thickness, we found that the simulation results showed good agreement with both Mode 1 and Mode 2 correlations, even for the high $Re$ and pressurized conditions.
	This fact suggests that if one can carefully treat the thermal-boundary-layer thickness, the forced-convection boiling HTC is expected to be accurately predicted with the present method.}

\section{Concluding remarks \label{sec:conclusions}}

In this paper, we developed a numerical approach based on the pseudopotential LBM to simulate forced-convection boiling.
To enhance numerical stability even at high Reynolds numbers, a CMs-based formulation was constructed.
The forcing scheme, consistent with the CMs-based LBM, led to a concise yet robust algorithm.
Furthermore, additional terms required to ensure thermodynamic consistency were derived in a CMs framework.
Four types of numerical tests verified the accuracy and convergence of the present approach. 

We then applied our CMs-based LBM to the system of convection-boiling heat transfer on a cylinder with the Reynolds number at 30 and 30000, respectively. 
Our forced-convection boiling simulations were not limited to the film boiling; a series of boiling processes, {\it i.e.}, nucleation, growth, departure of vapor bubbles, was also observed from simulations without any artificial input for phase change.
We demonstrated that 
the present scheme can reproduce all the boiling regimes. Moreover, the present scheme was able to well reproduce the characteristic behavior of the boiling curve even though the focused system was not the pool boiling but the forced-convection boiling system.
The simulation results supported the previous experimental observations of intermittent direct solid-liquid contact even in the film-boiling regime.
The obtained heat flux was evaluated on the ${Nu}$-$Ja$ diagram to compare with the semi-empirical correlations for the film boiling around a cylinder.
We found that our simulation results fall in between existing correlations,
but it was difficult to identify the best and unified correlations for different flow conditions.
As a result of improving the thermal boundary layer thickness, we found that the simulation results showed good agreement with both Mode 1 and Mode 2 correlations, even for the high $Re$ and pressurized conditions.
This fact suggests that if one can carefully treat with the thermal-boundary-layer thickness, the forced-convection boiling HTC is expected to be accurately predicted with the present method.

It is noted that Son and Dhir~\citep{Son2008} carried out pool boiling simulations and found a three-dimensional bubble-release pattern, which was confirmed by the experiment of Linehard and Dhir~\citep{Linehard1973}. 
Moreover, for the forced-convection system, the experiments using spheres by Liu and Theofanous~\citep{Liu1994} showed three-dimensional interface behaviors. 
On the other hand, Epstein and Hauser~\citep{Epstein1980} believed that differences between a cylinder and a sphere could be ignored in theoretical analysis. 
We think it would be necessary to perform further three-dimensional simulations and clarify these issues.


Our findings, however, provide improved understanding of forced-convection boiling on a cylinder.
The authors believe that the present CMs-based LBM is applicable to a wide range of  the high-Reynolds-number multiphase flows with phase change phenomena, which will be further exploited in follow-on research.

\section*{Acknowledgment}
	This work is supported by Mitsubishi Heavy Industries, Ltd.
S.S., Y.A., and A.K. are grateful to Mr. Hiroshi Sakaba and Mr. Hiroyuki Sato for their support.
S.S. greatly appreciates Mr. Kazuya Koyama's dedicated support over the years.
The support of JSPS KAKENHI Grant Number 16J02077 is also acknowledged.
Numerical simulations were performed on Oak Forest PACS (University of Tsukuba) and Earth Simulator (JAMSTEC).
A.D.R would like to thank Dr. Christophe Coreixas for insightful discussions and very valuable suggestions.
K.H.L. would like to acknowledge support from the UK Engineering and Physical Sciences Research Council under the project ``UK Consortium on Mesoscale Engineering Sciences (UKCOMES)'' (Grant Nos. EP/R029598/1 EP/P022243/1).

\section*{Data availability}
The data that support the findings of this study are available from the corresponding author
upon reasonable request.

\appendix

\section{Theoretical analysis\label{sec:Analysis}}

\begin{widetext}
	The Chapman--Enskog analysis for the present lattice Boltzmann model is provided here.
	We focus on the second-order analysis.
	For convenience of the analysis, we now rewrite the LBE in a general multiple-relaxation-time (GMRT) framework~\citep{Fei2017,Fei2018a}.
	To this end, we first decompose the matrix $\bm{T}$ into the transformation matrix $\bm{M}$ and the shift matrix $\bm{N}$~\citep{Fei2017}. 
	The transformation matrix $\bm{M}$ can be computed as
	\begin{equation}
	\bm{M} =\left [
	\begin{array}{c}
	\bra{|\boldsymbol{c}_i|^0} \\
	\bra{c_{ix}} \\
	\bra{c_{iy}} \\
	\bra{c_{ix}^2+ c_{iy}^2} \\
	\bra{c_{ix}^2- c_{iy}^2} \\
	\bra{c_{ix} c_{iy}}  \\
	\bra{c_{ix}^2 c_{iy}} \\
	\bra{c_{ix} c_{iy}^2}  \\
	\bra{c_{ix}^2 c_{iy}^2} 
	\end{array}
	\right] 
	= 
	\left [
	\begin{array}{ccccccccc}
	1& 1&  1&  1&  1& 1&  1&  1&  1 \\	
	0& 1& -1&  0&  0& 1& -1&  1& -1 \\	
	0& 0&  0&  1& -1& 1& -1& -1&  1 \\	
	0& 1&  1&  1&  1& 2&  2&  2&  2 \\	
	0& 1&  1& -1& -1& 0&  0&  0&  0 \\	
	0& 0&  0&  0&  0& 1&  1& -1& -1 \\ 
	0& 0&  0&  0&  0& 1& -1& -1&  1 \\ 
	0& 0&  0&  0&  0& 1& -1&  1& -1 \\ 
	0& 0&  0&  0&  0& 1&  1&  1&  1 \\ 
	\end{array}
	\right].
	\label{eq:transD2Q9}
	\end{equation}	
	The shift matrix can be given through the relation $\bm{T} = \bm{N} \bm{M}$:
	\begin{align}
	\bm{N} &= \bm{T} \bm{M}^{-1} \\
	& = 
	\left [
	\begin{array}{ccccccccc}
	1&0&0& 0& 0& 0&0&0& 0\\	
	-u_x&1&0& 0& 0& 0&0&0& 0\\	
	-u_y&0&1& 0& 0& 0&0&0& 0\\	
	u_x^2 + u_y^2& -2u_x& -2u_y& 1& 0& 0&0&0& 0\\	
	u_x^2 - u_y^2& -2u_x&2u_y& 0& 1& 0&0&0& 0\\	
	u_xu_y& -u_y& -u_x& 0& 0& 1&0&0& 0\\ 
	-u_x^2u_y&2u_xu_y&u_x^2&-u_y/2&-u_y/2&-2u_x&1&0& 0\\ 
	-u_xu_y^2&u_y^2&2u_xu_y&-u_x/2& u_x/2&-2u_y&0&1& 0\\ 
	u_x^2u_y^2& -2u_xu_y^2& -2u_x^2u_y& (u_x^2 + u_y^2)/2& -(u_x^2-u_y^2)/2& 4u_xu_y& -2u_y& -2u_x& 1\\ 
	\end{array}
	\right].		
	\label{eq:shiftD2Q9}
	\end{align}	
	The transformation matrix $\bm{M}$ transforms the distribution functions into the raw moments. 
	The shift matrix $\bm{N}$ transforms the raw moments into the central moments, and is a lower-triangular matrix.

	Using Eqs.~(\ref{eq:transD2Q9}) and (\ref{eq:shiftD2Q9}), we rewrite the LBE with forcing term as
	\begin{equation}
	f_i(\boldsymbol{x}+\boldsymbol{c}_i\delta_t,t+\delta_t) = f_i(\boldsymbol{x},t) - \Lambda_{ij}[f_j(\boldsymbol{x},t)-f^{\rm{eq}}_j(\boldsymbol{x},t)]
	+ \frac{\delta_t}{2} [F_i(\boldsymbol{x},t) + F_i(\boldsymbol{x}+\boldsymbol{c}_i\delta_t,t+\delta_t)],
	\label{eq:LBEbase}
	\end{equation}
	where ${\bm \Lambda} = \bm{M}^{-1}\bm{N}^{-1}\bm{K}\bm{N}\bm{M}$.
	Note that when we introduce the relation $\bar{f}_i=f_i-\delta_t F_i/2$, the implicitness of Eq.~(\ref{eq:LBEbase}) can be eliminated, and Eq.~(\ref{lbe}) in the manuscript can be recovered.
	For generality, we set the relaxation matrix as $\bm{K} = \mathrm{diag}[s_0,s_1,\dots,s_8] $ in this analysis.

	The Taylor-series expansion of Eq.~(\ref{eq:LBEbase}) at $({\boldsymbol{x}},t)$ yields:
	\begin{equation}
	\delta_t (\partial_t + \boldsymbol{c}_i \cdot \nabla)f_i 
	+ \frac{\delta_t^2}{2}(\partial_t + \boldsymbol{c}_i \cdot \nabla)^2 f_i
	= - \Lambda_{ij}[f_j-f_j^{\rm{eq}}]
	+ \delta_t F_i
	+ \frac{\delta_t^2}{2}(\partial_t + \boldsymbol{c}_i \cdot \nabla) F_i,
	\label{eq:with2ndOrder}
	\end{equation}	
	where we have neglected the $O(\delta_t^3)$ terms.
	Multiplying Eq.~(\ref{eq:with2ndOrder}) by $\bm{M}$ leads to
	\begin{equation}
	(\bm{I}\partial_t + \bm{D}) \boldsymbol{m}
	+ \frac{\delta_t}{2}(\bm{I}\partial_t + \bm{D})^2 \boldsymbol{m}
	= - \frac{ \bm{N}^{-1}\bm{K}\bm{N} }{\delta_t} (\boldsymbol{m}-\boldsymbol{m}^{\rm{eq}})
	+ \boldsymbol{S}
	+ \frac{\delta_t}{2}(\bm{I}\partial_t + \bm{D}) \boldsymbol{S},
	\end{equation}	
	where $\boldsymbol{m} = \bm{M} \ket{f_i}$, $\boldsymbol{m}^{\mathrm{eq}} = \bm{M} \ket{f_i^{\mathrm{eq}}}$, $\boldsymbol{S} = \bm{M} \ket{F_i}$, $\bm{D} = \bm{M} [( \boldsymbol{c}_i \cdot \nabla) \bm{I}] \bm{M}^{-1}$.
	To perform the Chapman--Enskog analysis, the following multiscale expansions are introduced:
	\begin{equation}
	\partial_t = \varepsilon \partial_{t_1} + \varepsilon^2 \partial_{t_2},~
	\nabla = \varepsilon \nabla_1,~
	f_i = f_i^{\rm eq} + \varepsilon f_{i}^{(1)} + \varepsilon^2 f_{i}^{(2)},~
	\boldsymbol{F} = \varepsilon \boldsymbol{F}^{(1)},
	\end{equation}
	which indicates that
	\begin{equation}
	\bm{D} = \varepsilon \bm{D}_1,~
	\boldsymbol{m} = \boldsymbol{m}^{\rm eq} + \varepsilon \boldsymbol{m}^{(1)} + \varepsilon^2 \boldsymbol{m}^{(2)},~
	\boldsymbol{S} = \varepsilon \boldsymbol{S}^{(1)},
	\end{equation}
	We can obtain:
	\begin{align}
	O(\varepsilon):& \, (\bm{I}\partial_{t_1} + \bm{D}_1) \boldsymbol{m}^{\rm eq}
	= - \frac{\bm{N}^{-1}\bm{K}\bm{N}}{\delta_t}\boldsymbol{m}^{(1)} 
	+ \boldsymbol{S}^{(1)}, 
	\label{eq:1stOrderLevel}\\
	O(\varepsilon^2):& \, \partial_{t_2} \boldsymbol{m}^{\rm eq} 
	+ \left(\bm{I}\partial_{t_1} + \bm{D}_1 \right) 
	\left( \bm{I} - \frac{
		\bm{N}^{-1}\bm{K}\bm{N}
	}{2} \right)\boldsymbol{m}^{(1)} =  - \frac{
		\bm{N}^{-1}\bm{K}\bm{N}
	}{\delta_t}\boldsymbol{m}^{(2)} .
	\label{eq:2ndOrderLevel}
	\end{align}		
	Here, we require the explicit expressions for $\bm{D}$, and $\bm{N}^{-1}\bm{K}\bm{N}$ to proceed the analysis.
	It is interesting to find the Chapman--Enskog analysis is basically the same with the one for non-orthogonal MRT models ({\it e.g.}, Refs.~\citep{Li2018b, Fei2019}), because $\bm{N}^{-1}\bm{K}\bm{N}$ itself is a {\it lower-triangular} matrix.

	From Eq.~(\ref{eq:1stOrderLevel}), we have the continuity and momentum equations at $O(\varepsilon)$ level:
	\begin{align}
	\partial_{t_1}\rho + \partial_{x_1}(\rho u_x) + \partial_{y_1}(\rho u_y) = 0, \label{eq:massOrder1} \\ 
	\partial_{t_1}(\rho u_x) + \partial_{x_1}\left(\frac{1}{3}\rho + \rho u_x^2 \right) + \partial_{y_1}(\rho u_xu_y) = F_x, \label{eq:momentumXOrder1} \\
	\partial_{t_1}(\rho u_y) + \partial_{x_1}(\rho u_xu_y) + \partial_{y_1} \left(\frac{1}{3}\rho  + \rho u_y^2 \right) = F_y. \label{eq:momentumYOrder1} 
	\end{align}
	Analogously, we have the ones at $O(\varepsilon^2)$ level:
	\begin{align}
	\partial_{t_2}\rho =& 0, \label{eq:massOrder2} \\ 
	\partial_{t_2}(\rho u_x) + \frac{1}{2} \partial_{x_1} \left[ \left(1-\frac{s_3}{2} \right) m_3^{(1)}
	+ \left(1-\frac{s_4}{2} \right)m_4^{(1)} \right]
	+ \partial_{y_1} \left[ \left(1-\frac{s_5}{2} \right)m_5^{(1)} \right] =& 0, \label{eq:momentum_x} \\
	\partial_{t_2}(\rho u_y) 
	+ \partial_{x_1} \left[ \left(1-\frac{s_5}{2} \right)m_5^{(1)} \right]
	+ \frac{1}{2} \partial_{y_1} \left[ \left(1-\frac{s_3}{2} \right) m_3^{(1)}
	- \left(1-\frac{s_4}{2} \right)m_4^{(1)} \right] =& 0,  \label{eq:momentum_y}
	\end{align}
	Here we should specify $m_3^{(1)}$, $m_4^{(1)}$, $m_5^{(1)}$.
	By using the $O(\varepsilon)$ equation again, we have 
	\begin{align}
	\partial_{t_1} m_3^{\mathrm{eq}} 
	+ \partial_{x_1} (m_1^{\mathrm{eq}} + m_7^{\mathrm{eq}})
	+ \partial_{y_1} (m_2^{\mathrm{eq}} + m_6^{\mathrm{eq}}) =
	-s_3 \frac{m_{3}^{(1)}}{\delta_t} + S_3^{(1)}, \label{eq:m3}\\
	\partial_{t_1} m_4^{\mathrm{eq}} 
	+ \partial_{x_1} (m_1^{\mathrm{eq}} - m_7^{\mathrm{eq}})
	- \partial_{y_1} (m_2^{\mathrm{eq}} - m_6^{\mathrm{eq}}) =
	-s_4 \frac{m_{4}^{(1)}}{\delta_t} + S_4^{(1)}, \\
	\partial_{t_1} m_5^{\mathrm{eq}} 
	+ \partial_{x_1} m_6^{\mathrm{eq}} 
	+ \partial_{y_1} m_7^{\mathrm{eq}}  =
	-s_5 \frac{m_{5}^{(1)}}{\delta_t} + S_5^{(1)}. \label{eq:m5}
	\end{align}
	Eqs.~(\ref{eq:m3})--(\ref{eq:m5}) can be rewritten as 
	\begin{align}
	m_3^{(1)} = & - \frac{2\delta_t}{s_3} \rho c_s^2 (\partial_{x_1} u_x + \partial_{y_1} u_y ) 
	\underline{\underline{+ \frac{2\sigma |\boldsymbol{F}_m|^2}{(1-s_3/2) \psi^2} }}
	\underline{+(u_x^3 \partial_{x_1} \rho 
		+u_y^3\partial_{y_1} \rho 
		+3 \rho u_x^2 \partial_{x_1} u_x
		+3 \rho u_y^2 \partial_{y_1} u_y)}, \label{eq:mm3}\\
	m_4^{(1)} = & 	- \frac{2\delta_t}{s_4} \rho c_s^2(\partial_{x_1}u_x - \partial_{y_1}u_y) 
	\underline{+ (u_x^3\partial_{x_1} \rho - u_y^3 \partial_{y_1} \rho
		+3 \rho u_x^2 \partial_{x_1} u_x - 3 \rho u_y^2 \partial_{y_1} u_y)}, \label{eq:mm4}\\
	m_5^{(1)} = & -\frac{\delta_t}{s_5} \rho c_s^2 (\partial_{x_1}u_y + \partial_{y_1}u_x),
	\end{align}
	where we set $c_s^2 = 1/3$.
	Under low Mach-number assumptions, the underlined terms in Eqs.~(\ref{eq:mm3}) and (\ref{eq:mm4}) may be neglected.
	Recently, Huang {\it et al.}~\citep{Huang2018} derived correction terms to eliminate such third-order terms. 
	The double underlined term in Eq.~(\ref{eq:mm3}) have appeared due to the modification described in Sec.~\ref{sec:formulation}.
	Then, Eqs.~(\ref{eq:momentum_x}) and (\ref{eq:momentum_y}) becomes
	\begin{align}
	\partial_{t_2}(\rho u_x) = & 
	\partial_{x_1} \left[\rho \nu (2\partial_{x_1}{u_x}) 
	+ \rho (\nu_b-\nu) (\nabla \cdot \boldsymbol{u})
	- \frac{\sigma |\boldsymbol{F}_m|^2}{\psi^2} 
	\right] 
	+ \partial_{y_1} \left[ \rho \nu (\partial_{x_1} u_y + \partial_{y_1} u_x)  \right], \label{eq:momentumXorder2} \\
	\partial_{t_2}(\rho u_y) = & 
	\partial_{x_1} \left[\rho \nu (\partial_{x_1} u_y + \partial_{y_1} u_x) 
	\right]
	+ \partial_{y_1} \left[
	\rho \nu (2\partial_{y_1}{u_y} )
	+\rho (\nu_b-\nu) (\nabla \cdot \boldsymbol{u})
	- \frac{\sigma |\boldsymbol{F}_m|^2}{\psi^2}
	\right] \label{eq:momentumYorder2} ,
	\end{align}
	where $\nu_b = c_s^2 \delta_t(\omega_b^{-1}-1/2)$ and $\nu = c_s^2 \delta_t(\omega^{-1}-1/2)$ with $s_3=\omega_b$ and $s_4=s_5=\omega$, respectively.
	
	Combining Eqs.~(\ref{eq:massOrder1}) with (\ref{eq:massOrder2}) through $\partial_t = \partial_{t_1} + \varepsilon \partial_{t_2}$, the continuity equation can be obtained
	\begin{equation}
	\partial_t \rho + \nabla \cdot (\rho \boldsymbol{u}) = 0.
	\end{equation}	
	Similarly, combining Eqs.~(\ref{eq:momentumXOrder1})--(\ref{eq:momentumYOrder1}) with Eqs.~(\ref{eq:momentumXorder2})--(\ref{eq:momentumYorder2}), we can obtain the Navier--Stokes equation as follows:
	\begin{equation}
	\partial_t (\rho \boldsymbol{u}) 
	+ \nabla \cdot (\rho \boldsymbol{u} \boldsymbol{u}) = 
	- \nabla \cdot (\rho c_s^2 + \sigma G^2 c_s^4 |\nabla \psi|^2)\bm{I}
	+ \nabla \cdot [\rho \nu (\nabla \boldsymbol{u} + \nabla \boldsymbol{u}^\top) + \rho(\nu_b - \nu)(\nabla \cdot \boldsymbol{u})  \bm{I} \,] + \boldsymbol{F} .
	\label{eq:modifiedNS}
	\end{equation}	
	We can find that the pressure tensor has been modified to be thermodynamically consistent, which is similar to the scheme proposed for the orthogonal MRT model~\citep{Li2013b}.
	
\end{widetext}

%
%

\section{Unit conversion of heat flux\label{sec:Conversion}}

Here, we convert several quantities in lattice units into physical units through appropriate dimensionless parameters.
As for the physical properties in physical units, we use the values at $T = 556 $ K (the saturation temperature of the present paper).

First, we convert the cylinder diameter $D$ based on the capillary length $l_c$ as follows:
\begin{equation}
D^{p.u.} = \left(\frac{D^{l.u.}}{l_c^{l.u.}} \right) l_c^{p.u.},
\end{equation}	
where the upper indices ``$l.u.$'' and ``$p.u.$'' are abbreviations for ``lattice units'' and ``physical units'', respectively.
The capillary length $l_c$ can be given by
\begin{equation}
l_c = \left(\frac{\gamma}{\Delta \rho g}  \right)^{1/2},
\end{equation}	
where $\gamma$, $\Delta \rho=\rho_l-\rho_v$, and $g$ are the surface tension coefficient, the density difference, and the gravitational acceleration, respectively.
With the parameter set used in this paper (including the surface tension coefficient measured from the Laplace test in Sec.~\ref{sec:tests}) and physical properties at $T=556$ K available from the steam table~\citep{JSME}, we can calculate the cylinder diameter in physical units as $D^{p.u.}=0.93$ mm. 
Since we set the cylinder diameter in lattice units to $D^{l.u.}=30$ as in Sec.~\ref{sec:simulation}, the corresponding mesh size can be evaluated as $\Delta x = 0.93/30 = 0.031$ mm.

The characteristic velocity $u^{p.u.}$ is determined via the Reynolds number:
\begin{equation}
u^{p.u.} = \left( \frac{\nu_l^{p.u.}}{D^{p.u.}} \right) Re.
\end{equation}
In this paper, we set two types of Reynolds numbers: $Re=30$ and $30000$.
The velocity in physical units $u^{p.u.}$ can be calculated as $4.01\times10^{-3}$ m/s and $4.01$ m/s for $Re=30$ and $30000$, respectively.
The time step size $\Delta t$ can be determined from the dimensionless time scale $t^*=t/(D/u)$;
$\Delta t$ in physical units can be written by
\begin{equation}
\Delta t^{p.u.} = \frac{D^{p.u.}/u^{p.u.}}{D^{l.u.}/u^{l.u.}} \Delta t^{l.u.},
\label{eq:realTimeStep}
\end{equation}
where $\Delta t^{l.u.}=\delta_t$ is usually set to 1 in the LBM.
With Eq.~(\ref{eq:realTimeStep}), one can calculate the time step sizes for $Re=30$ and $30000$ as $\Delta t = 0.387$ ms and $0.000387$ ms, respectively.

The wall superheat degree  $\Delta T$ can be scaled by the critical temperature as follows:
\begin{equation}
\Delta T^{p.u.} = \frac{\Delta T^{l.u.}}{T_c^{l.u.}}T_c^{p.u.}.
\end{equation}

Finally, we convert the heat flux into physical units.
The heat flux for the forced-convection system can be non-dimensionalized by~\citep{Katto1980}
\begin{equation}
q^* = \frac{q}{G h_{fg}},
\label{eq:nonDimHeatFlux}
\end{equation}
where $G=\rho_l u$ is the mass velocity of the flowing liquid.
With Eq.~(\ref{eq:nonDimHeatFlux}), we can convert the heat flux into physical units as follows:
\begin{equation}
q^{p.u.} = \frac{G^{p.u.} h_{fg}^{p.u.}}{G^{l.u.} h_{fg}^{l.u.}} q^{l.u.}.
\end{equation}

\section{Film boiling HTC correlations\label{sec:correlations}}

\subsection{Bromley et al. correlation}
The original equation derived by Bromley {\it et al.}~\citep{Bromley1953} is inconvenient because the derivative appears non-linearly. 
After a certain amount of {\it ad hoc} approximation, the equation results in a simple form.
At high flow regime ($u_l/\sqrt{gD}>2$), Bromley {\it et al.}'s correlation is given by
\begin{equation}\label{eq:Bromley}
{Nu} = C \,
\eta^{1/2}
Re^{1/2}	 Pr^{1/2} \left(\frac{c_{p,v}\Delta T}{h_{fg}'} \right)^{-1/2} ,
\end{equation}
where	$C=2.7$ and $h_{fg}' = h_{fg} (1+ 0.4c_{p, v}\Delta T/h_{fg} )^2$ is the effective latent heat of vaporization~\citep{Bromley1952}.

\subsection{Epstein--Hauser correlation}
\citet{Epstein1980} originally derived the HTC correlation for subcooled forced-convection film boiling.
When the subcooling degree is assumed to be zero, their correlation degrades into
\begin{equation} \label{eq:Epstein}
{Nu} =  C\, \gamma^{1/4}
\eta^{1/2}
Re^{1/2}
Pr^{1/4}  
{Ja}^{-1/4}, 
\end{equation}
where $C=0.553$ for sphere and $0.537$ for cylinder from theoretical analysis; 
we used $C=0.537$ in this paper since we focused on the heat transfer on a cylinder.
Note that they finally concluded that  $C=1.13$, used in this paper,  provided a reasonable correlation of  observed HTC for  the experimental data of  subcooled forced-convection film boiling from spheres or cylinders~\citep{Bromley1950,Bromley1953,Motte1957,Dhir1978}. 

\subsection{Ito et al. correlation}
Ito {\it et al.}~\citep{Ito1981} analyzed the forced-convection film boiling heat transfer from a horizontal cylinder to saturated liquid based on the two-phase boundary layer theory.
For the conditions of predominant forced convection, they obtained the equation identical to Eq.~(\ref{eq:Epstein}).
They obtained the constants as $C=0.46$, $0.48$, and $0.51$ for water, ethanol, and hexane, respectively; we used $C=0.46$ in this paper.

\section{Bubble growth simulation\label{sec:bubbleGrowth}}

\begin{figure}
	\centering
	\includegraphics[width=0.65\linewidth]{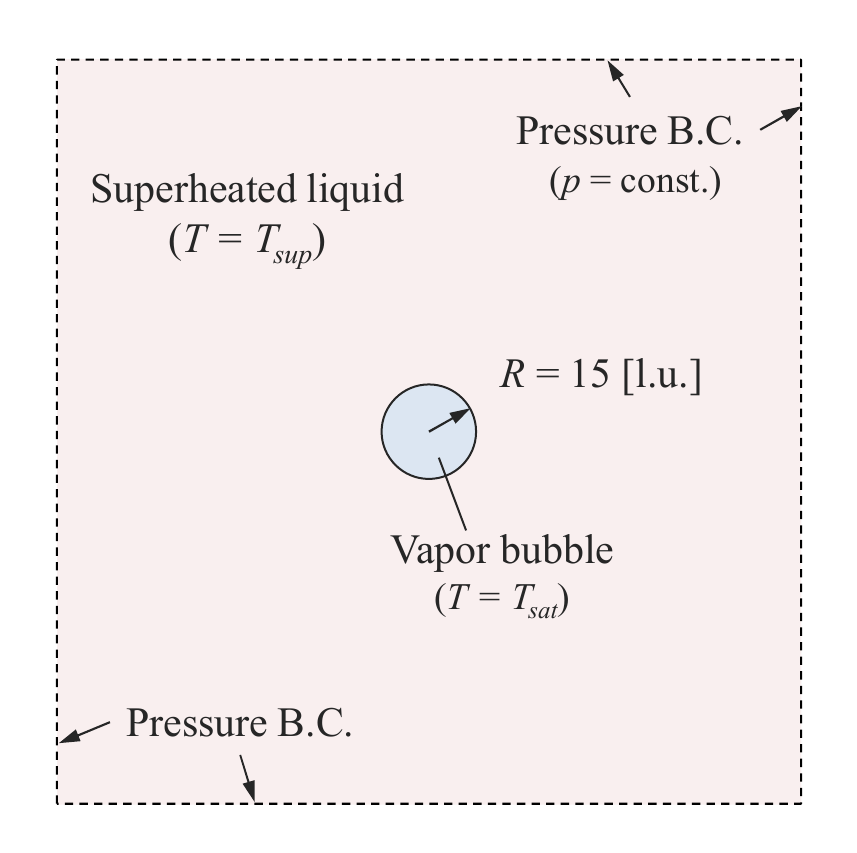}
	\caption{
		Computational setup for the simulation of bubble growth in an overheated liquid.
	}
	\label{fig:figure20}
\end{figure}

To understand the effect of the thermal diffusivity of the liquid phase on the thermal boundary layer,
the simulations of bubble growth in an overheated liquid are carried out.
The results are compared with the Plesset--Zwick Eq.~\citep{Plesset1954}, which is given by
\begin{equation}\label{eq:Plesset}
R(t) = \left(\frac{12}{\pi} \alpha_l \right)^{1/2} 
\frac{\rho_l c_{p,l} \Delta T}{\rho_v h_{fg}} t^{1/2},
\end{equation}
where $r$ is the bubble diameter, $t$ is the time, and $\alpha_l$ is the thermal diffusivity for the liquid phase.

Fig.~\ref{fig:figure20} shows the schematic diagram of the numerical setup. 
The computational domain is discretized into $1000\times1000$.
The initial bubble radius and superheat degree are set to $R(0) = 15$ and $\Delta T = 0.04T_c$, respectively.

\begin{figure*}[tb]
	\centering
	\includegraphics[width=0.9\linewidth]{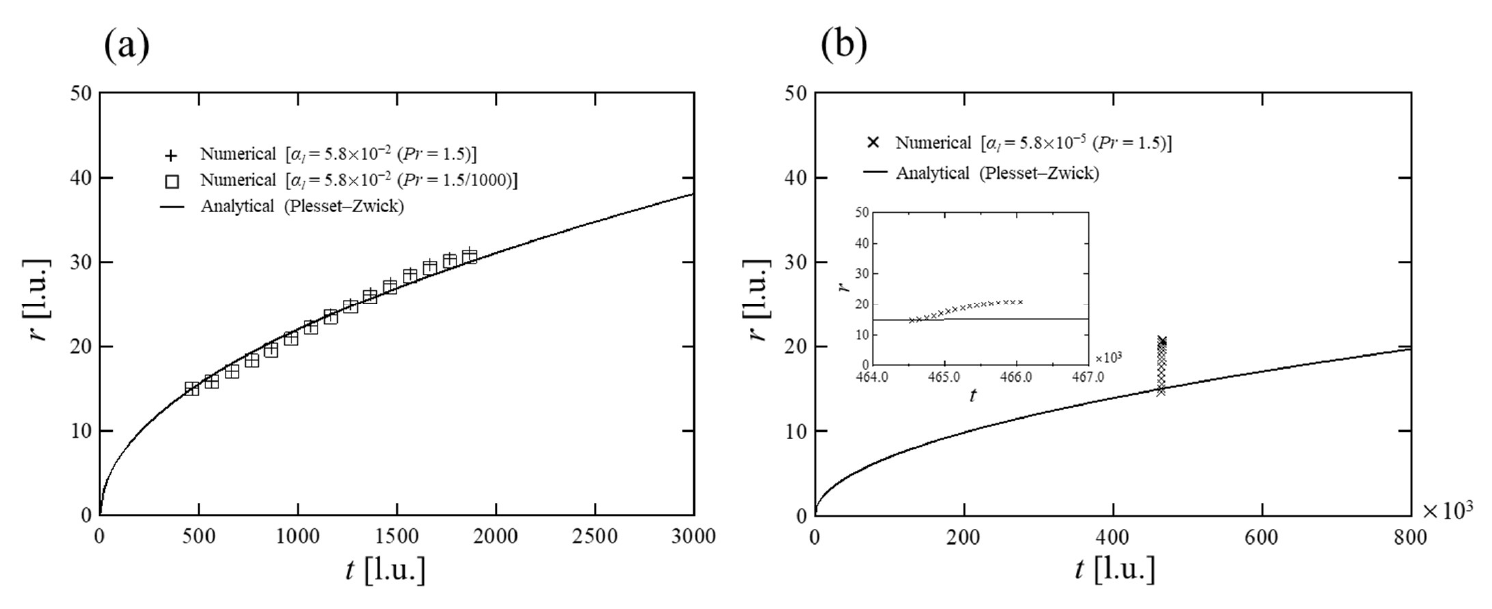}
	\caption{
		Comparison of the present simulation results with the Plesset--Zwick equation: 
		(a) $\alpha_l = 5.8\times 10^{-2}$, (b) $\alpha_l = 5.8\times10^{-5}$ in lattice units.
	}
	\label{fig:figure21}
\end{figure*}

The simulation results are shown in Fig.~\ref{fig:figure21}. 
The thermal diffusivity and kinematic viscosity for the liquid phase (or Prandtl number) used here correspond to the ones employed in the simulations of Sec.~\ref{sec:simulation}. 
We can find that the high-$\alpha_l$ case [Fig.~\ref{fig:figure21}(a)] well agree with the analytical solution,
while the low-$\alpha_l$ case [Fig.~\ref{fig:figure21}(b)] disagrees with the analytical one. 
This is considered  to be influenced by the thermal boundary layer. 
For the latter case, the spatial resolution is insufficient to capture the thermal-boundary-layer thickness.

\clearpage
\section*{References}
\bibliography{references}

\end{document}